\documentclass[sigplan,10pt]{acmart}
\usepackage[normalem]{ulem}
\usepackage{blindtext}
\setcopyright{acmcopyright}
\ccsdesc[500]{Computer systems organization~Distributed architectures}
\ccsdesc[300]{Computer systems organization~Cloud computing}
\ccsdesc[300]{Computing methodologies~Distributed artificial intelligence}
\acmYear{2026}\copyrightyear{2026}
\setcopyright{cc}
\setcctype[4.0]{by}
\acmConference[EUROSYS '26]{European Conference on Computer Systems}{April 27--30, 2026}{Edinburgh, Scotland Uk}
\acmBooktitle{European Conference on Computer Systems (EUROSYS '26), April 27--30, 2026, Edinburgh, Scotland Uk}
\acmDOI{10.1145/3767295.3769316}
\acmISBN{979-8-4007-2212-7/26/04}

\usepackage[colorinlistoftodos]{todonotes}
\usepackage{algorithmic}
\usepackage{caption}
\usepackage{subcaption} 
\usepackage{fancybox}
\usepackage{xcolor}
\usepackage{array} 
\usepackage{booktabs} 
\usepackage{makecell}
\usepackage{multirow}
\usepackage{multicol} 
\usepackage{tabularx} 
\usepackage{graphicx}
\usepackage{amsmath}
\usepackage{amsfonts}
\usepackage{comment}
\usepackage{colortbl}
\usepackage{multicol}
\usepackage{array}
\usepackage{hyperref}
\usepackage{pifont}
\usepackage{textcomp}
\usepackage[linesnumbered,ruled,vlined]{algorithm2e}
\usepackage{enumitem}
\usepackage{framed}
\usepackage{color}
\setlist[itemize]{noitemsep,leftmargin=*}
\SetKw{Continue}{continue}
\usepackage{listings}
\usepackage{titlesec}

\hypersetup{
  colorlinks=true,
  linkcolor=violet,
  filecolor=magenta,      
  urlcolor=cyan,
}
\titlespacing*{\section}{0pt}{10pt}{8pt}
\titlespacing*{\subsection}{0pt}{9pt}{6pt}

\newcommand{\tp}[1]{\noindent \textbf{#1}\hspace{0.5em}}
\newcommand{\ts}[1]{\noindent \textbf{\textit{#1}}}

\newcommand{\figref}[1]{\textcolor{violet}{Fig.~\ref{#1}}}
\newcommand{\Sref}[1]{\textcolor{violet}{\S\ref{#1}}}
\newcommand{\eqtref}[1]{\textcolor{violet}{Eq.~\ref{#1}}}
\newcommand{\tableref}[1]{\textcolor{violet}{Table~\ref{#1}}}

\newcommand{\Sys}{{\scshape FlexPipe}\xspace}

\begin{document}
\title{\Sys: Adapting Dynamic LLM Serving \\ Through Inflight Pipeline Refactoring in Fragmented Serverless Clusters}

\author{Yanying Lin}
\affiliation{
    \institution{Shenzhen Institutes of Advanced Technology, CAS; UCAS}
    \country{}
}
\affiliation{
    \institution{UC San Diego}
    \country{}
}

\author{Shijie Peng}
\affiliation{
    \institution{Shenzhen Institutes of Advanced Technology, CAS}
    \country{}
}
\affiliation{
    \institution{UCAS}
    \country{}
}

\author{Chengzhi Lu}
\affiliation{
    \institution{Nanyang Technological University}
    \country{}
}

\author{Chengzhong Xu}
\affiliation{
    \institution{University of Macau}
    \country{}
}

\author{Kejiang Ye\textsuperscript{*}}
\affiliation{
    \institution{Shenzhen Institutes of Advanced Technology, CAS}
    \country{}
    \authornote{Corresponding Author}
}
\renewcommand{\shortauthors}{Yanying Lin, Shijie Peng, Chengzhi Lu, Chengzhong Xu, and Kejiang Ye}
\begin{abstract}
    Serving Large Language Models (LLMs) in production faces significant challenges from highly variable request patterns and severe resource fragmentation in serverless clusters. Current systems rely on static pipeline configurations that struggle to adapt to dynamic workload conditions, leading to substantial inefficiencies. 
    
    We present \Sys, a novel system that dynamically reconfigures pipeline architectures during runtime to address these fundamental limitations. \Sys decomposes models into fine-grained stages and intelligently adjusts pipeline granularity based on real-time request pattern analysis, implementing three key innovations: fine-grained model partitioning with preserved computational graph constraints, inflight pipeline refactoring with consistent cache transitions, and topology-aware resource allocation that navigates GPU fragmentation. Comprehensive evaluation on an 82-GPU cluster demonstrates that \Sys achieves up to 8.5× better resource efficiency while maintaining 38.3\% lower latency compared to state-of-the-art systems, reducing GPU reservation requirements from 75\% to 30\% of peak capacity.
\end{abstract}

\keywords{Large Language Models, Serverless Computing, Resource Efficiency, Pipeline Parallelism, Dynamic Scaling}

\maketitle

\section{Introduction} \label{sec:introduction}
The exponential growth in Large Language Model (LLM) parameters \cite{touvronLLaMAOpenEfficient2023,touvronLlamaOpenFoundation2023,brownLanguageModelsAre2020,devlinBERTPretrainingDeep2019,anilPaLMTechnicalReport2023} has created significant challenges for serving these models in production environments. With models scaling to hundreds of billions of parameters, deploying efficient inference systems requires distributed computing approaches as single-device execution becomes infeasible due to memory constraints \cite{aminabadiDeepSpeedInferenceEnabling2022, narayananPipeDreamGeneralizedPipeline2019, zhengAlpaAutomatingInter2022}. Current serving systems primarily rely on two distributed paradigms: tensor parallelism, which distributes matrix operations across devices with high-bandwidth interconnects, and pipeline parallelism, which segments models into sequential stages with lower communication requirements \cite{huangGPipeEfficientTraining2019, zhuohanli0001AlpaServeStatisticalMultiplexing2023}. However, these approaches face substantial challenges when deployed in real-world serverless environments \cite{yangINFlessNativeServerless2022, linQUARTLatencyAwareFaaS2024}, particularly in balancing computational efficiency with resource adaptability under dynamic request patterns \cite{hongzhang0025SHEPHERDServingDNNs2023} and fragmented cluster resources \cite{qizhenwengBewareFragmentationScheduling2023}.

\begin{figure}[]
    \begin{minipage}[]{.33\linewidth}
        \centering
        \centerline{\includegraphics[width=\linewidth]{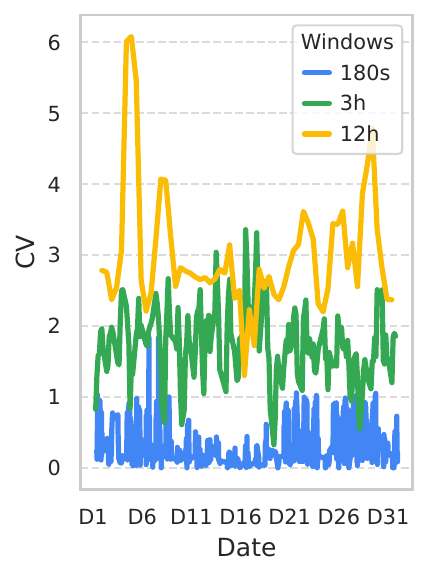}}
        \vspace{-0.4em}
        \subcaption{Alibaba Trace}\medskip
        \label{sfig:cv_results_lora}
    \end{minipage}
    \hspace{-0.5em}
    \begin{minipage}[]{.33\linewidth}
        \centering
        \centerline{\includegraphics[width=\linewidth]{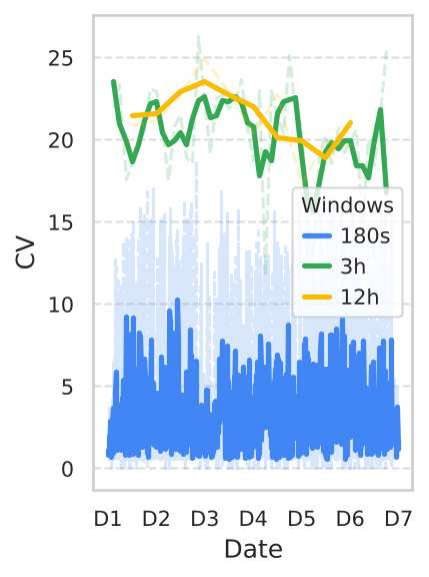}}
        \vspace{-0.4em}
        \subcaption{Top-1 App}\medskip
        \label{sfig:cv_results_0insggn2}
    \end{minipage}
    \begin{minipage}[]{.33\linewidth}
        \centering
        \centerline{\includegraphics[width=\linewidth]{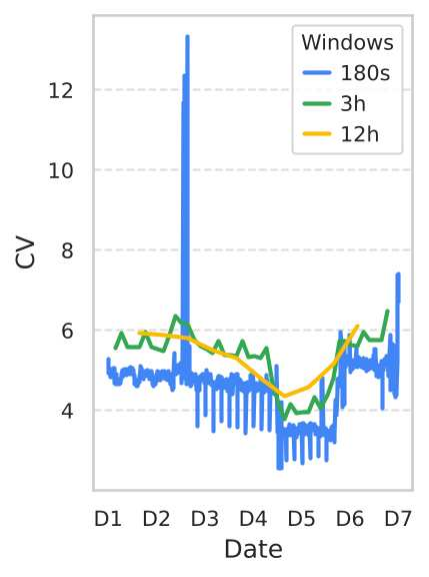}}
        \vspace{-0.4em}
        \subcaption{Top-2 App}\medskip
        \label{sfig:cv_results_n0nh88df}
    \end{minipage}
    \vspace{-1.2em}
    \caption{Request distribution CV (coefficient of variation) variations across different periods. Significant mismatches exist in CV calculated with different window sizes (180s, 3h, 12h), 7$\times$ variation exists. (a) Request distribution CV of Alibaba Trace, (b) Request distribution CV of Top-1 App, and (c) Top-2 App from Azure \cite{zhangFasterCheaperServerless2021}.}
    \label{fig:cv_dist} 
    \vspace{-1.5em}
\end{figure}

Production analysis of Alibaba clusters \footnote{Request distribution traces and GPU data are openly available at \url{https://github.com/alibaba/clusterdata/tree/master/cluster-trace-v2026-GenAI}.} reveals \textit{two fundamental challenges} that current LLM serving systems cannot address. \textbf{i) workload volatility}: request patterns exhibit extreme variability with coefficient of variation (CV) fluctuating up to 7× across timeframes (\figref{fig:cv_dist}), causing static pipelines to misalign with shifting workload characteristics \cite{hongzhang0025SHEPHERDServingDNNs2023, gyeong-inyuOrcaDistributedServing2022} and resulting in 17\% average GPU utilization. \textbf{ii) resource fragmentation}: serverless environments scatter GPUs across heterogeneous workloads \cite{qizhenwengBewareFragmentationScheduling2023, jieliTetrisMemoryefficientServerless2022}, preventing the high-bandwidth interconnects essential for tensor parallelism \cite{zhengAlpaAutomatingInter2022, aminabadiDeepSpeedInferenceEnabling2022}. Our measurements show only 0.02\% probability of co-locating 4 GPUs on the same server (\Sref{sec:motivation}), forcing suboptimal execution patterns. This architectural mismatch between model requirements and fragmented resource availability fundamentally undermines serving efficiency in elastic environments \cite{yangINFlessNativeServerless2022, linQUARTLatencyAwareFaaS2024}.

Existing systems \cite{zhuohanli0001AlpaServeStatisticalMultiplexing2023,liDeepSpeedDataEfficiency2023, kwonEfficientMemoryManagement2023,hongzhang0025SHEPHERDServingDNNs2023, duanMuxServeFlexibleSpatialTemporal2024,agrawalTamingThroughputLatencyTradeoff2024,huInferenceInterferenceDisaggregate2024} employ sophisticated pipeline optimization techniques but fundamentally rely on static configurations that cannot adapt to dynamic environments. While these systems achieve impressive performance under stable conditions, they struggle with the dual challenges of workload volatility and resource fragmentation. For instance, AlpaServe \cite{zhuohanli0001AlpaServeStatisticalMultiplexing2023} optimizes pipeline architectures based on historical request patterns, focusing on long-term performance rather than adapting to short-term variability. Such static approaches inevitably create bottlenecks when faced with rapid workload fluctuations or when fragmented GPU resources prevent optimal tensor-parallel execution (\Sref{sec:motivation}). This mismatch between fixed pipeline designs and the dynamic reality of serverless environments results in significant inefficiencies during deployment.

To address these fundamental limitations, we introduce \Sys, a dynamically adaptive LLM serving system that performs inflight pipeline refactoring without service interruption. \Sys challenges the conventional wisdom that fixed pipeline configurations are necessary for consistent performance. Instead, it exploits a critical insight: pipeline granularity requirements fundamentally shift with workload characteristics. Fine-grained pipelines excel under bursty workloads by distributing buffering capacity across stages, while coarse-grained pipelines minimize communication overhead during stable periods. \Sys continuously transitions between these configurations based on real-time coefficient of variation metrics, achieving superior resource efficiency across the full spectrum of serverless request patterns.

However, implementing this approach presents three significant technical challenges: (1) determining optimal partition boundaries that balance computation and communication overhead while preserving computational graph constraints for future reconfiguration, (2) maintaining cache consistency during dynamic pipeline topology transitions without service interruption, and (3) navigating GPU resource fragmentation in highly dynamic serverless environments while minimizing cold-start initialization latency.

To address these challenges, \Sys introduces three core innovations: (1) \textit{Fine-grained model partitioning} that decomposes LLMs through a constrained optimization algorithm, balancing computation and communication overhead while preserving computational graph constraints for efficient future reconfiguration; (2) \textit{Inflight pipeline refactoring} that dynamically transitions between pipeline granularities without service interruption, using real-time monitoring and CV metrics to seamlessly reconfigure pipeline topologies while maintaining cache consistency; and (3) \textit{Topology-aware resource allocation} that employs hierarchical resource coordination and memory-aware scheduling strategies to navigate GPU fragmentation, minimizing contention during parallel scaling operations while transforming cold starts into efficient warm starts through parameter locality preservation.

We evaluated our approach on a production-grade Kubernetes cluster with 42 servers and 82 GPUs using realistic workloads. Results demonstrate significant performance advantages: 38.3\% lower latency under stable workloads and 66.1\% improvement under variable conditions. For large models like OPT-66B, \Sys achieves 24.38\% lower prefill latency compared to alternatives. Most importantly, our system maintains consistent performance as workload variability increases—recovering from pipeline stalls in just 9ms under high-CV conditions (82\% faster than competitors) while achieving up to 8.5× better resource efficiency. In a production deployment, our dynamic resource allocation strategy reduced always-on GPU reservation from 75\% to just 30\% of peak capacity without compromising service quality, while decreasing resource allocation wait time by 85\%. These results confirm that \Sys effectively addresses the fundamental challenges of resource fragmentation and request variability in serverless environments.

\tp{Contributions.} Our key contributions include:
\begin{itemize}[noitemsep]
    \item A novel approach for dynamically reconfiguring pipeline architectures in response to changing request distributions without service interruption.
    \item A method for fine-grained model partitioning that enables efficient pipeline refactoring while preserving computational efficiency.
    \item A system for enhancing LLM inference elasticity through dynamic resource allocation and pipeline topology adaptation.
    \item Empirical evidence demonstrating \Sys's effectiveness through extensive evaluation on production-grade infrastructure with real-world workloads.
\end{itemize}

\section{Background}\label{background}
\subsection{Distributed LLM Inference Paradigms}
The exponential growth in LLM parameters has created fundamental conflicts between memory capacity and computational demands. With models scaling to hundreds of billions of parameters, single-device approaches face severe memory constraints, while inference scenarios encounter compute limitations under concurrent requests \cite{patelSplitwiseEfficientGenerative2024,zhangFasterCheaperServerless2021,shahradServerlessWildCharacterizing2020,wengMLaaSWildWorkload2022}. Distributed parallel computing has emerged as the essential solution for large-scale model deployment \cite{grattafioriLlama3Herd2024, touvronLlamaOpenFoundation2023}.

\ts{Tensor Parallelism:} Tensor parallelism distributes tensor operations across multiple devices by partitioning matrix operations row-wise or column-wise \cite{aminabadiDeepSpeedInferenceEnabling2022, zhengAlpaAutomatingInter2022}. This approach effectively parallelizes multi-head attention mechanisms in Transformers and achieves optimal computational efficiency when GPUs are tightly coupled with high-bandwidth interconnects. However, tensor parallelism fundamentally depends on low-latency, high-bandwidth network communication (NVLink, InfiniBand) for frequent synchronization operations \cite{narayananPipeDreamGeneralizedPipeline2019}. This network dependency becomes a critical bottleneck in fragmented environments where GPUs are distributed across different physical nodes with limited inter-node bandwidth. In such scenarios, the frequent all-reduce operations required for tensor synchronization can dominate execution time, making tensor parallelism impractical for distributed deployments with commodity network infrastructure.

\ts{Pipeline Parallelism:} Pipeline parallelism implements an inter-layer decoupling strategy, dividing models into sequential stages based on layer dependencies \cite{huangGPipeEfficientTraining2019, athlurVarunaScalableLowcost2022a, rajbhandariZeROMemoryOptimizations2020, aminabadiDeepSpeedInferenceEnabling2022, fanDAPPLEPipelinedData2021}. This approach employs asynchronous scheduling to achieve spatiotemporal overlap between computation and communication \cite{narayananPipeDreamGeneralizedPipeline2019, zhuohanli0001AlpaServeStatisticalMultiplexing2023}. The fundamental advantage of pipeline parallelism lies in its communication pattern: while tensor parallelism requires $O(n^2)$ all-reduce communications per layer with $n$ devices, pipeline parallelism reduces inter-stage communication to point-to-point transfers with $O(1)$ complexity per stage. This dramatic reduction in communication overhead—from dense synchronization matrices to sparse sequential dependencies—enables pipeline parallelism to maintain performance even when network bandwidth is constrained or latency is high. 
The asynchronous nature of pipeline execution also enables natural load balancing across heterogeneous hardware configurations, as slower devices can process smaller pipeline stages without blocking faster ones. However, this communication efficiency comes with the inherent challenge of pipeline bubble overhead, where stages remain idle during pipeline fill and drain phases. Sophisticated micro-batching strategies and overlapping techniques are essential to amortize these bubbles and maintain high computational efficiency \cite{huangGPipeEfficientTraining2019, liDeepSpeedDataEfficiency2023}.


\subsection{Serverless Challenges for LLM Serving}
Serverless computing architectures promise enhanced hardware utilization through dynamic resource provisioning \cite{zengMedusaAcceleratingServerless2025,zilizhangJolteonUnleashingPromise2024,sahraeiXFaaSHyperscaleLow2023,schmidSeBSFlowBenchmarkingServerless2025}. However, the fundamental design philosophy of serverless platforms creates a fundamental tension with distributed LLM inference requirements.

Modern serverless schedulers \cite{zilizhangJolteonUnleashingPromise2024,fuServerlessLLMLowLatencyServerless2024} implement anti-affinity policies that deliberately scatter service replicas across diverse physical nodes to prvent cascading failures. This spatial distribution conflicts directly with distributed LLM inference, which requires tightly coupled GPU clusters with high-bandwidth interconnects (e.g., NVLink, InfiniBand) for efficient tensor parallelism. The result is a paradoxical scenario: individual GPUs are abundant, but cohesive GPU clusters are scarce, fundamentally undermining communication-intensive parallelism strategies.

The serverless resource allocation model operates through dual-tier provisioning: \textit{always-on resources} (60-75\% of peak capacity) guarantee baseline service levels, while \textit{elastic resources} handle demand spikes. This conservative approach, designed to prevent service outages, creates chronic underutilization during normal operations while still introducing multi-second scaling delays that violate sub-second response requirements for interactive LLM applications.

Beyond serverless, similar fragmentation challenges emerge in multi-tenant clusters enforcing strict isolation, edge computing deployments with heterogeneous hardware, and dedicated clusters supporting diverse workloads. 
The common thread across these environments is the fundamental tension between resource isolation policies designed for predictable performance and the collaborative access patterns required for efficient distributed inference.

This analysis reveals that resource fragmentation represents a fundamental systems challenge requiring adaptive pipeline architectures that maintain efficiency across diverse and changing resource landscapes, rather than static optimization approaches that assume stable resource topologies.

\section{Motivation}\label{sec:motivation}
Optimizing parallel computing strategies is essential for efficient LLM inference, particularly in serverless environments with fragmented resources and fluctuating request patterns. Our systematic analysis reveals critical correlations between inference performance, pipeline architecture, and request distributions that fundamentally impact serving efficiency.

\subsection{Resource Fragmentation in Cloud}

\begin{table}[]
    \centering
    \caption{GPU cluster statistics showing resource utilization patterns.}
    \vspace{-1em}
    \footnotesize
    \renewcommand{\arraystretch}{1.2}
    \setlength{\tabcolsep}{5pt}
    \begin{tabularx}{0.47\textwidth}{l>{\raggedleft\arraybackslash}X>{\raggedleft\arraybackslash}X}
        \toprule
        \rowcolor{gray!10} \textbf{Metric} & \textbf{Cluster C1} & \textbf{Cluster C2} \\
        \midrule
        Number of nodes & 430 & 927 \\
        \rowcolor{gray!5} Number of GPUs & 468 & 1,175 \\
        \midrule
        \multicolumn{3}{l}{\textit{SM Utilization (\%)}} \\
        Mean & 16.91 & 23.74 \\
        \rowcolor{gray!5} Median (P50) & 9.16 & 10.85 \\
        P95 & 80.53 & 85.37 \\
        \rowcolor{gray!5} 10-30\% utilization & 31.26\% & 20.98\% \\
        \midrule
        \multicolumn{3}{l}{\textit{GPU Memory Utilization (\%)}} \\
        Mean & 43.48 & 50.92 \\
        \rowcolor{gray!5} Median (P50) & 28.78 & 53.69 \\
        P95 & 99.09 & 99.34 \\
        \rowcolor{gray!5} 10-30\% utilization & 38.44\% & 17.78\% \\
        \bottomrule
    \end{tabularx}
    \vspace{0.2em}
    \caption*{\footnotesize{ \textbf{Note:} C1 represents the inference-only cluster, while C2 is a hybrid training-inference cluster. Both employ dynamic resource scaling strategies.}}
    \label{tab:cluster_stats}
    \vspace{-2.5em}
\end{table}

\begin{figure}[]
    \begin{minipage}[]{.48\linewidth}
        \centering
        \centerline{\includegraphics[width=\linewidth]{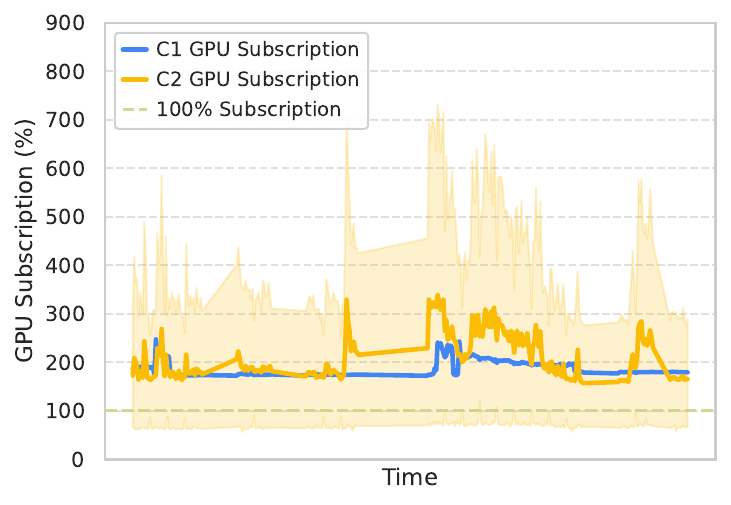}}
        \vspace{-0.2em}
        \subcaption{GPU Subscription Rate}\medskip
        \label{sfig:GPU_Subscription_Comparison}
    \end{minipage}
    \hfill
    \begin{minipage}[]{.48\linewidth}
        \centering
        \centerline{\includegraphics[width=\linewidth]{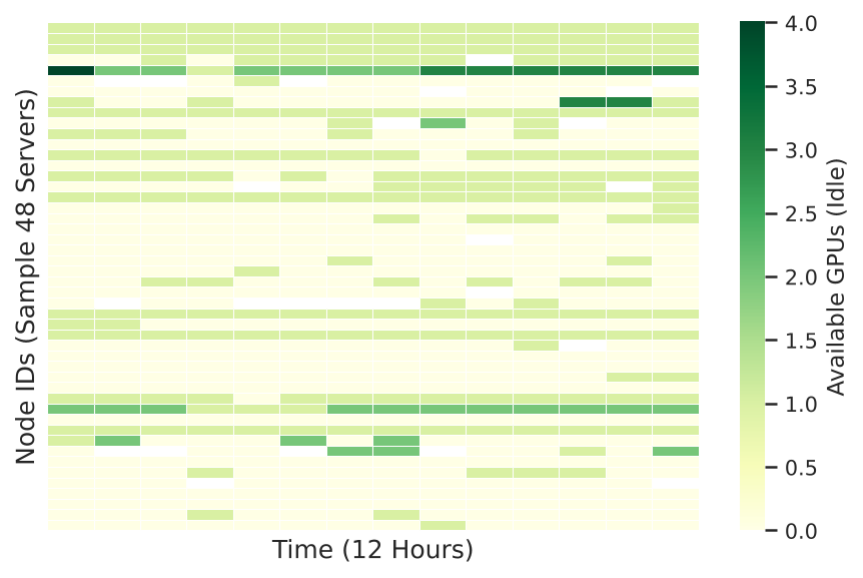}}
        \vspace{-0.2em}
        \subcaption{GPU Availability Heatmap}\medskip
        \label{sfig:gpu_availability_heatmap}
    \end{minipage}
    \vspace{-1.em}
    \caption{Resource fragmentation in Alibaba. (a) GPU subscription rate averaging 216\%, indicating significant resource overcommitment, and (b) Heatmap revealing spatially scattered GPU availability patterns that impede formation of high-bandwidth interconnected GPU groups needed for tensor parallelism.}
    \label{fig:fragmentation} 
    \vspace{-1em}
\end{figure}

To understand resource characteristics in cloud environments, we conducted a two-week analysis of GPU resources from a major cloud provider. Our findings revealed a striking 216\% average GPU subscription rate (\figref{sfig:GPU_Subscription_Comparison}), indicating that \textit{two services typically share each GPU}. Memory utilization presented significant variability (\tableref{tab:cluster_stats}), with P50 servers showing modest 20.3\% GPU memory utilization while P99 servers approached saturation at 99.3\%.

Resource availability analysis demonstrates severe constraints: securing a single GPU with >85\% free memory occurs with only 8.7\% probability, while co-locating 4 GPUs on the same server drops to 0.02\%. This fragmentation stems from heterogeneous model deployments creating unpredictable memory patterns, performance isolation mechanisms fragmenting resources, and aggressive oversubscription maximizing utilization at the cost of availability.

The practical impact of this fragmentation is substantial. In production clusters, 78\% of tensor parallelism requests were forced to degrade to pipeline parallelism due to unavailability of adjacent GPUs with high-bandwidth interconnects. This degradation fundamentally challenges distributed inference paradigms, as tensor operations designed for tightly-coupled execution must be reorganized into pipeline stages with higher communication overhead and suboptimal memory access patterns. Communication-intensive approaches like tensor parallelism require high-bandwidth interconnections between GPUs, yet the \textit{scattered distribution of available GPUs} (\figref{sfig:gpu_availability_heatmap}) creates persistent misalignment between physical topology and logical requirements. The \textit{ephemeral nature} of serverless GPU allocation further exacerbates this challenge, as optimal GPU configurations may remain available only briefly before reallocation—creating an inherent incompatibility between tensor-parallel computational models and the reality of fragmented cloud GPU environments.

This GPU fragmentation introduces \textit{significant operational challenges}. Due to the immediate reallocation of released GPUs to competing workloads, production clusters typically adopts conservative scaling strategies—maintaining approximately 75\% of historical peak GPU capacity as always-on resources, with the remaining 25\% allocated through dynamic scaling. This approach produces a \textbf{problematic trade-off}: during normal operations, GPU utilization remains unnecessarily low (approximately 17\% in our measurement study), yet during traffic spikes, the delayed provisioning of additional GPUs frequently causes SLO violations as scaling operations cannot keep pace with request bursts. The \textit{fundamental disconnect} between idealized theoretical GPU allocation models and the reality of fragmented, ephemeral GPU availability creates a persistent efficiency gap in large-scale LLM serving.

\ts{Insight 1:}\textit{
    Resource fragmentation in cloud environments significantly impedes communication-intensive parallelism strategies that rely on high-bandwidth interconnects, necessitating alternative approaches for distributed LLM inference.
}

\subsection{Pipeline Granularity and Data Parallelism} \label{sec:loading}
Pipeline parallelism provides an effective solution for utilizing fragmented GPU resources in LLM inference. The granularity of pipeline stages—defined by operators or parameters per stage—fundamentally affects memory footprint, communication patterns, and computational characteristics. As shown in \tableref{tab:performance_metrics}, finer-grained partitioning significantly reduces per-stage memory requirements, decreasing parameter loading time and per-stage inference latency, but introduces a critical trade-off: more stages cause proportionally increased inter-stage communication overhead.

\begin{table}[]
    \centering
    \vspace{-1em}
    \footnotesize
    \caption{Performance metrics for different pipeline granularities.}
    \vspace{-1em}
    \label{tab:performance_metrics}
    \renewcommand{\arraystretch}{1.3}
    \setlength{\tabcolsep}{5pt}
    \begin{tabularx}{0.47\textwidth}{c|cccc}
    \toprule
    \rowcolor{gray!10} \textbf{Stages} & \textbf{Load(s)} & \textbf{Compute(ms)} & \textbf{Comm.(ms)} & \textbf{Max Batch} \\
    \midrule
    4  & 47.14 & 69.94 &  6.3  & 128  \\
    \rowcolor{gray!5}
    8  & 13.05 & 36.63 & 14.7  & 256  \\
    16 &  9.19 & 18.67 & 31.5  & 512  \\
    \rowcolor{gray!5}
    32 &  5.43 &  9.67 & 65.1  & 1024 \\
    \bottomrule
    \end{tabularx}
    \footnotesize{\textit{Note: OPT-66B (120GB) performance with sequence length 4096 on A100 GPUs. \textbf{Compute} indicates per-stage inference time, \textbf{Comm.} represents inter-stage communication overhead, and \textbf{Max Batch} shows maximum supported batch size per configuration.}}
    \vspace{-2.5em}
\end{table}

In elastic serverless environments, fine-grained pipeline architectures offer significant advantages. Our experiments show 32-stage pipelines reduce parameter loading latency to just 5.43s—an 8.7× improvement over 4-stage configurations—enabling rapid establishment of data-parallel replicas during demand spikes. This reduction incurs a 65.1ms communication penalty per inference iteration, creating a trade-off between initialization speed and runtime efficiency.

Fine-grained partitioning creates a fundamental trade-off in memory efficiency: 32-stage pipelines achieve 8× larger batch sizes (1024 vs 128) than 4-stage configurations for OPT-66B, dramatically improving GPU tensor core utilization. This increased batch capacity amortizes communication overhead across more requests, creating a counterintuitive effect where higher communication costs are offset by improved computational efficiency.

This granularity trade-off reveals a fundamental architectural insight that directly motivates dynamic adaptation in serverless environments. The dichotomy between fine-grained and coarse-grained pipeline configurations exposes an inherent temporal optimization problem: bursty serverless workloads demand fine-grained pipelines to achieve rapid horizontal scaling (8.7× faster initialization) and exploit large batch processing capacity (8× larger batches) during traffic spikes, while stable operational periods derive greater efficiency from coarser partitions that minimize per-request communication overhead through reduced inter-stage coordination. This creates a dynamic optimization landscape where the optimal pipeline configuration is fundamentally time-dependent and workload-sensitive. The strategic imperative becomes clear: systems must dynamically transition between these configurations—temporarily adopting fine-grained architectures during demand surges to maximize elasticity and batch throughput, then reverting to coarser-grained configurations during stable periods to minimize communication penalties and optimize per-request latency.

Given the inherently bursty nature of serverless workloads, dynamic pipeline granularity adjustment becomes essential. During traffic spikes, fine-grained pipelines provide \textit{rapid scaling} and increased batch processing capacity. As traffic stabilizes, transitioning to coarser-grained configurations minimizes per-request latency through reduced communication overhead, creating an optimal balance between elasticity and efficiency.

\ts{Insight 2:}\textit{
    Fine-grained pipelines offer superior elasticity and batch processing capability during bursty workloads but incur communication overhead penalties. The optimal approach dynamically transitions between granularities—using fine-grained configurations temporarily during traffic spikes and reverting to coarser pipelines when workloads stabilize.
}

\subsection{Pipeline Overhead in Request Distributions} \label{sec:cv}
Advanced pipeline systems~\cite{zhuohanli0001AlpaServeStatisticalMultiplexing2023,aminabadiDeepSpeedInferenceEnabling2022,agrawalTamingThroughputLatencyTradeoff2024,huInferenceInterferenceDisaggregate2024} primarily optimize pipeline architectures offline using historical workload data for long-term performance. However, LLM inference in serverless environments demands short-term optimization: cloud providers need to quickly reclaim idle GPU resources, while multi-model services and multi-agent systems require locally optimized performance through concentrated resources. This fundamental mismatch between long-term optimization strategies and short-term load characteristics leads to significant performance degradation, manifested as pipeline bubbles and queuing delays.

\begin{figure}[]
    \begin{minipage}[]{.33\linewidth}
        \centering
        \centerline{\includegraphics[width=\linewidth]{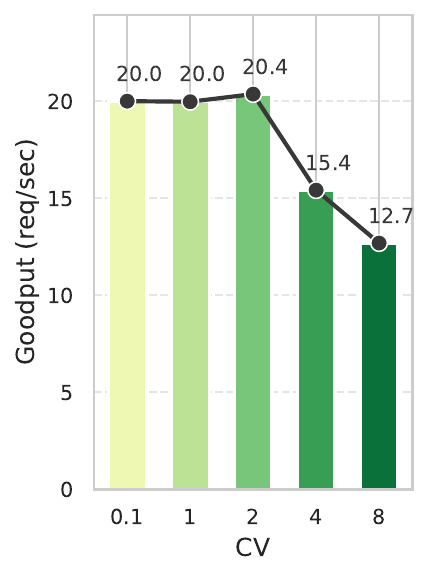}}
        \vspace{-0.2em}
        \subcaption{Goodput}\medskip
        \label{sfig:goodput_param_cv_comparison}
    \end{minipage}
    \begin{minipage}[]{.32\linewidth}
        \centering
        \centerline{\includegraphics[width=\linewidth]{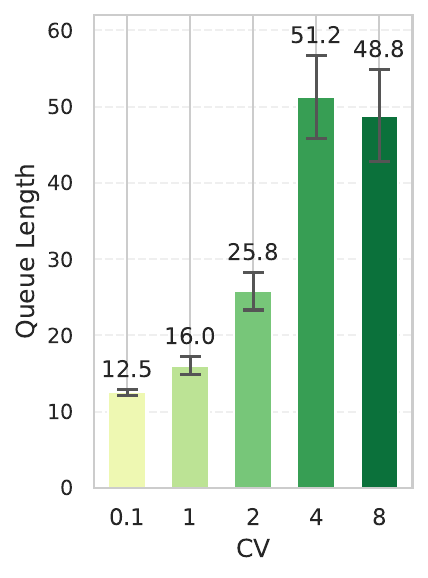}}
        \vspace{-0.2em}
        \subcaption{Queue Length}\medskip
        \label{sfig:queue_length_param_cv_comparison_bar}
    \end{minipage}
    \begin{minipage}[]{.33\linewidth}
        \centering
        \centerline{\includegraphics[width=\linewidth]{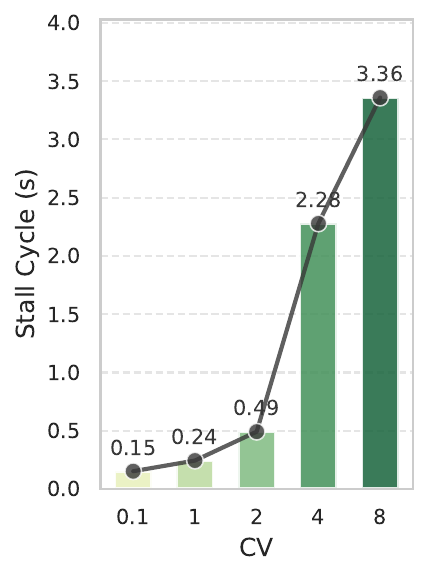}}
        \vspace{-0.2em}
        \subcaption{Stall Cycle}\medskip
        \label{sfig:stall_cycle}
    \end{minipage}
    \vspace{-1.5em}
    \caption{Impact of request distribution variability on pipeline performance. (a) \textit{Goodput} decreases by 37\% as CV increases from 0.1 to 8 due to resource contention; (b) Average \textit{queue length} grows nearly 4× with increasing CV, indicating pipeline congestion; (c) \textit{Stall cycle} ratio increases exponentially (22×) at high CV values, showing how static pipelines become inefficient under variable workloads.}
    \label{fig:stall} 
    \vspace{-1.em}
\end{figure}
As shown in \figref{fig:cv_dist}, request distribution CV exhibits \textit{substantial fluctuation}, with variation of up to \textbf{7×} across different time windows. To quantify how this volatility impacts performance, we evaluated a static 4-stage OPT-66B pipeline under varying request distributions with a baseline QPS of 20. \figref{fig:stall} reveals the \textit{severe performance degradation} caused by increasing workload variability across multiple dimensions. As CV increases from 0.1 to 8, goodput decreases by 37\% (\figref{sfig:goodput_param_cv_comparison}), while average queue length grows nearly \textbf{4×} (\figref{sfig:queue_length_param_cv_comparison_bar})—clear indicators of pipeline congestion and request backpressure. Most critically, the pipeline stall cycle ratio increases \textit{exponentially}, reaching \textbf{22×} at high CV values (\figref{sfig:stall_cycle}), demonstrating how static pipeline configurations become \textit{fundamentally inefficient} under variable workloads.

Our experiments demonstrate that with the same 4-stage pipeline architecture, a 4× difference in CV leads to a nearly 10× increase in pipeline stall overhead (comparing CV=1 and CV=4). This exponential relationship reveals the \textit{critical importance of adapting pipeline structures to match request volatility patterns}, as static configurations optimized for one CV value perform poorly when the request distribution changes.

\begin{figure}[]
    \begin{minipage}[]{.49\linewidth}
        \centering
        \centerline{\includegraphics[width=\linewidth]{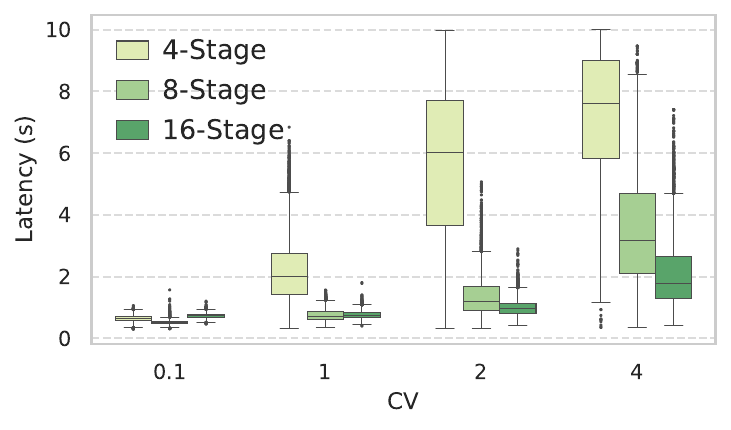}}
        \vspace{-0.2em}
        \subcaption{Dist. of Latency across CVs}\medskip
        \label{sfig:M1_box_delay}
    \end{minipage}
    \hfill
    \begin{minipage}[]{.49\linewidth}
        \centering
        \centerline{\includegraphics[width=\linewidth]{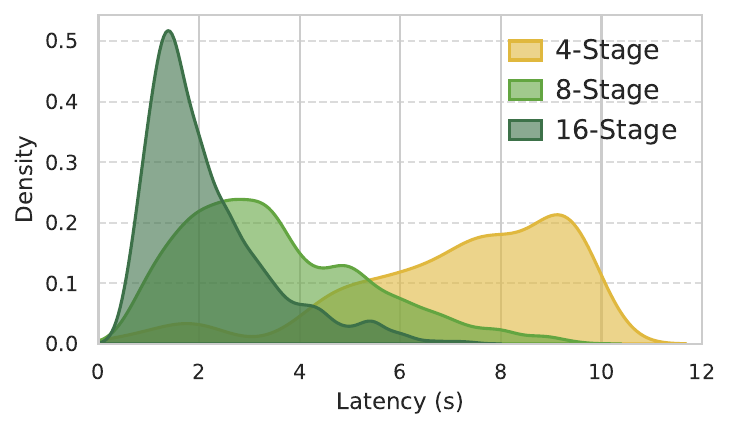}}
        \vspace{-0.2em}
        \subcaption{Latency Dist. of CV=4}\medskip   
        \label{sfig:M1_latency_distribution_cv4_stage_under3}
    \end{minipage}
    \vspace{-1.2em}
    \caption{Latency distribution across different request patterns. (a) Box plot comparing pipeline granularities across varying CV values, showing fine-grained pipelines perform better with high-variability workloads; (b) Detailed latency distribution for CV=4 with 4-stage pipeline, revealing significant variance from pipeline stalls.}
    \label{fig:cv} 
    \vspace{-1.em}
\end{figure}

To explore the relationship between pipeline architecture and load characteristics, we evaluated three pipeline models (4, 8, and 16 stages) under constant total request volume but varying CV values. As shown in \figref{fig:cv}, 4-stage and 8-stage architectures maintain excellent response time stability (approximately 0.5 seconds) under low CV conditions, while the 16-stage architecture's processing time increases to 1.2 seconds (2.7× longer). However, in high-burst scenarios (CV=4), the 16-stage architecture achieves average latency of only \textit{one-third} that of the 4-stage architecture, comparable to the latter's performance at CV=1.

This behavior can be explained through a stochastic process model that reveals the dynamic coupling between pipeline depth ($S$) and load burstiness ($CV$). While the theoretical delay of an $S$-stage pipeline is $T_{pipe} = S \cdot \tau + (S-1)\cdot\delta$ (where $\tau$ is single-stage service time and $\delta$ is communication overhead), burst requests cause uneven workloads across stages. We established an extended G/G/S queuing model:

\begin{equation}
    \footnotesize
    T_{total} = \underbrace{\frac{\rho^{S}}{S!(1-\rho)} \cdot \frac{CV_a^2 + CV_s^2}{2}}_{\text{Queue Latency}} + \underbrace{\sum_{i=1}^S \left( \frac{\lambda_i}{\mu_i - \lambda_i} \right)}_{\text{Stage Congestion Delay}}
\end{equation}
where $\rho = \lambda/\mu$ represents system utilization, $\lambda$ is arrival rate, $\mu$ is service rate, $CV_a$ and $CV_s$ denote coefficients of variation for arrival intervals and service times respectively, $\lambda_i$ is the arrival rate at stage $i$, and $\mu_i$ is the service rate at stage $i$. As $CV_a$ increases, increasing pipeline stages $S$ produces two opposing effects: 1) fine-grained task segmentation reduces per-stage service time, alleviating congestion; 2) it increases cumulative pipeline register delays. Our experimental data indicates that when $CV_a > 3$, effect 1 dominates, and setting $S \propto \sqrt{CV_a}$ achieves optimal latency. This explains why the 16-stage pipeline achieves a 3× performance improvement over the 4-stage pipeline at CV=4—the deeper pipeline \textit{effectively absorbs peak loads through distributed buffering}.

\ts{Insight 3:}\textit{
    Request distribution variability causes pipeline stage imbalances, resulting in different optimal architectures for various workloads. In highly bursty environments, deeper pipeline architectures can effectively absorb peak loads through distributed buffering, significantly outperforming static configurations.
}

\section{System Overview}
\begin{figure}[]
    \centering
    \includegraphics[width=0.96\linewidth]{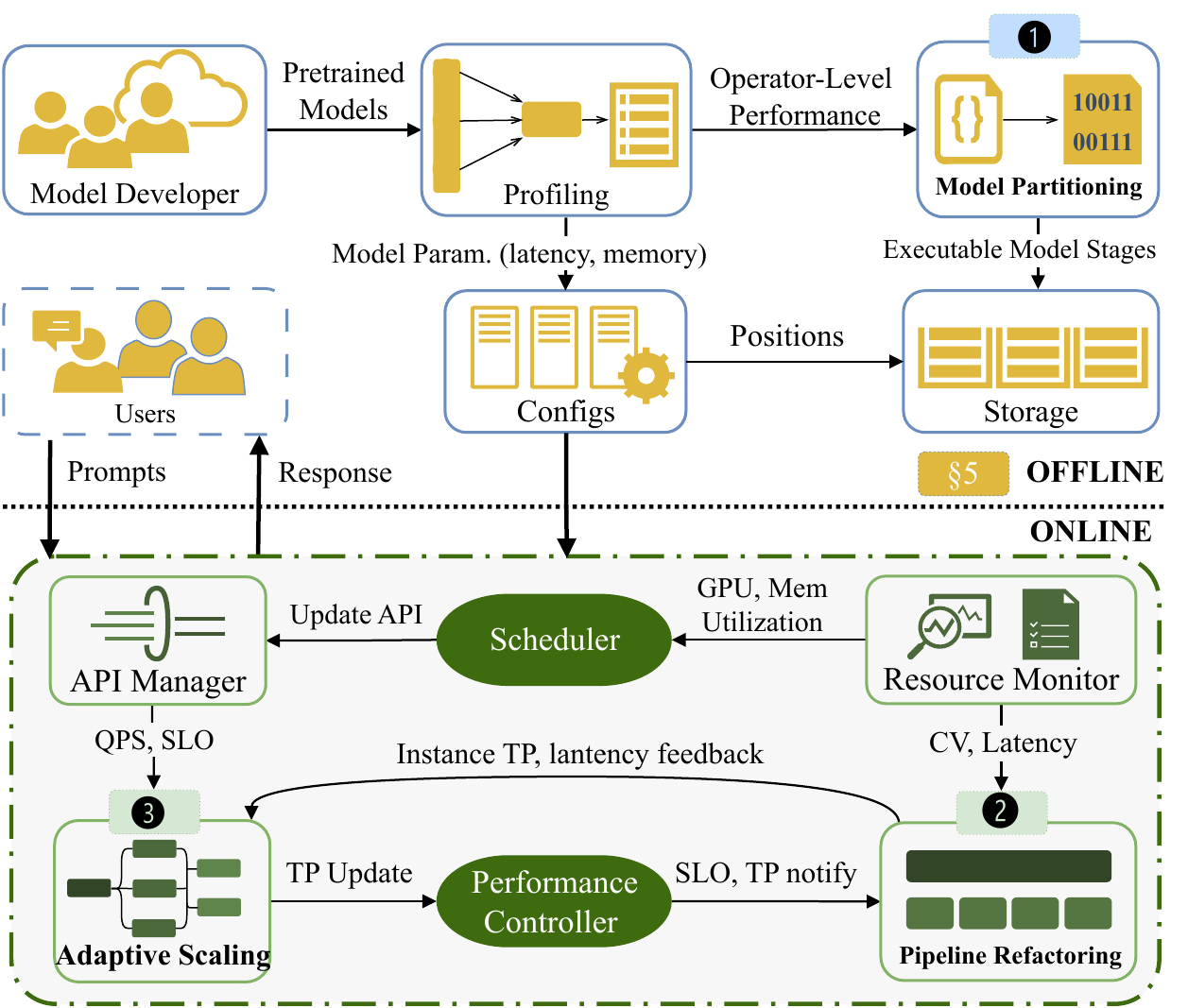}
    \vspace{-.5em}
    \caption{\Sys system architecture showing the three core components: \ding{182} Fine-Grained Pipeline Model Partitioning that decomposes LLMs at operator level for optimal adaptability, \ding{183} Inflight Pipeline Refactoring that dynamically adjusts pipeline granularity based on request patterns, and \ding{184} Adaptive Pipeline Scaling that enables efficient resource allocation during traffic fluctuations.}
    \label{fig:overview}
    \vspace{-2em}
\end{figure}

Based on our analysis of resource fragmentation, pipeline granularity trade-offs, and dynamic request patterns in serverless environments, we present \Sys, a dynamically adaptive LLM serving system designed to overcome these challenges. As illustrated in \figref{fig:overview}, \Sys comprises three synergistic components:

\textbf{Fine-Grained Pipeline Model Partitioning.} 
This component performs operator-level decomposition of LLMs to create balanced pipeline stages optimized for both computation and reconfiguration potential. By analyzing computation graphs and establishing natural partition boundaries, it creates pipeline architectures ranging from coarse stages with minimal communication overhead to fine-grained stages that enable rapid scaling during bursty workloads.

\textbf{Inflight Pipeline Refactoring.} 
At \Sys's core, this component monitors request distributions and dynamically restructures pipeline topology without interruption. Using coefficient of variation (CV) metrics and queue monitoring, it selects optimal configurations that minimize stalls while maximizing resource utilization. The system seamlessly transitions between fine-grained pipelines (for high-CV traffic) and coarse-grained architectures (for stable workloads) through consistent parameter migration.

\textbf{Adaptive Pipeline Scaling.} 
This component orchestrates GPU allocation during traffic fluctuations through topology-aware scheduling. It implements a Hierarchical Resource Graph to coordinate parallel scaling while avoiding resource contention, and employs affinity-based scheduling to leverage parameter locality across scaling events, transforming cold starts into efficient warm starts through intelligent cache management.

\ts{Key Implementation Challenges.} 
Implementing \Sys presents three critical technical challenges: (1) determining optimal operator-level partition boundaries that balance computation-communication trade-offs while preserving refactoring potential, requiring specialized constrained optimization algorithms; (2) maintaining state consistency during topology changes without service interruption through efficient KV cache synchronization and coordinated parameter migration; and (3) navigating resource fragmentation while minimizing initialization delays using topology-aware allocation strategies that preserve parameter locality across scaling operations. These challenges represent the fundamental tension between dynamic adaptation and efficient resource utilization in serverless environments.

\section{Fine-Grained Model Partitioning} \label{sec:offline_pipeline_optimization}
To enable adaptation to varying request patterns, \Sys decomposes models into fine-grained pipeline stages that can be reconfigured at runtime. The core challenge lies in balancing communication overhead and computational efficiency while facilitating dynamic refactoring. This requires solving three critical issues: determining the optimal granularity that satisfies bandwidth constraints while enabling communication-computation overlap, maintaining performance stability across varying micro-batch sizes, and supporting seamless runtime pipeline transitions with minimal synchronization overhead.

\Sys initiates pipeline optimization through computation graph analysis and operator-level profiling. For a given model $M$ with $L$ layers, we first construct its computation graph $G=(V,E)$ where vertices $v_i \in V$ represent operators and edges $e_{ij} \in E$ denote data dependencies. The Profiling module measures three critical metrics for each operator: computation time $t_c(v_i)$, parameter size $s_p(v_i)$, and activation size $s_a(v_i)$. To achieve optimal partitioning, we employ a dynamic programming algorithm that simultaneously considers communication-computation overlap and future refactoring needs.

The partitioning process solves a constrained optimization problem:
\begin{equation}
\begin{aligned}
    \footnotesize
\min_{\{S_k\}} & \sum_{k=1}^K \left| t_c(S_k) + \frac{s_p(S_k)}{B} - C \right| + \lambda \cdot R(S_k) \\
\text{s.t.} & \bigcup_{k=1}^K S_k = V, \quad S_i \cap S_j = \emptyset \ \forall i \neq j \\
& \max_{S_k} s_p(S_k) \leq M_{\text{GPU}} 
\end{aligned}
\end{equation}
where $K$ denotes the number of stages, $S_k$ represents the $k$-th stage containing a subset of operators, $V$ is the complete set of operators in the computation graph, $t_c(S_k)$ is the computation time of stage $k$, $s_p(S_k)$ is the parameter size of stage $k$, $B$ represents the inter-stage bandwidth, $C$ is the target computation-communication overlap cycle, $M_{\text{GPU}}$ is the GPU memory capacity, and $\lambda$ is a regularization weight. The regularization term $R(S_k)$ encodes the refactoring potential of each partition, favoring cuts that preserve hierarchical structure boundaries (e.g., attention blocks in Transformers) to facilitate future merging. This formulation ensures balanced stage execution times while creating natural breakpoints for potential pipeline reconfiguration.

For micro-batch adaptation, we introduce batch-aware transmission scaling:
\begin{equation}
s_a(S_k, b) = s_a^{base}(S_k) \cdot \left(1 + \alpha \log\frac{b}{b_{\text{base}}}\right)
\end{equation}
where $b$ is the micro-batch size, $b_{\text{base}}$ is the profiling batch size, and $\alpha$ is the compression factor learned from historical data through linear regression. This allows the system to predict communication patterns for arbitrary batch sizes during online serving. Crucially, the partitioning algorithm preserves the parameter grouping structure to enable future replica alignment - parameters within the same logical group (e.g., attention heads or MLP blocks) are colocated in contiguous memory regions, allowing merged stages to reuse existing memory layouts.

\section{Inflight Pipeline Refactoring} \label{sec:online_pipeline_optimization}
The dynamic nature of cloud environments and the resource fragmentation challenges identified earlier necessitate a flexible approach to pipeline management. To address this, we design an inflight pipeline refactoring mechanism (\figref{fig:variant}) that can dynamically adjust pipeline granularity during model serving without service interruption. The core challenge lies in dynamically adjusting pipeline granularity under time-varying request distributions while maintaining hardware efficiency and consistency.

We formulate this as a multi-objective optimization problem with temporal constraints. Let $\mathcal{G} = \{g_1,...,g_K\}$ denote the set of candidate pipeline granularities, where each granularity $g_k = (\eta_k, b_k)$ corresponds to stage count $\eta_k$ and batch size $b_k$. The temporal correlation of request patterns is captured by coefficient of variation (CV) $\nu_t = \frac{\sigma_t}{\mu_t}$, where $\sigma_t$ and $\mu_t$ represent the standard deviation and mean of request arrival intervals at time $t$, respectively.

\subsection{Granularity Adaptation}
The optimal granularity $g^*$ is determined through joint optimization of throughput and latency, balancing the trade-off between processing speed and response time:
\begin{equation}
    g^* = \arg\max_{g_k \in \mathcal{G}} \left[ \alpha \cdot \frac{T_k}{T_{max}} + (1-\alpha) \cdot \frac{L_{min}}{L_k} \right] \cdot \exp\left(-\frac{|\nu_t - \nu_k|}{\sigma}\right)
    \label{eq:optimal_granularity}
\end{equation}
where $g_k = (\eta_k, b_k)$ represents granularity configuration $k$ with stage count $\eta_k$ and batch size $b_k$, $\mathcal{G}$ is the set of candidate granularities, $T_k$ and $L_k$ denote throughput and latency for granularity $g_k$, $T_{max}$ and $L_{min}$ are normalization constants, $\nu_t$ is the current CV value at time $t$, $\nu_k$ represents the optimal CV threshold for granularity $g_k$, $\alpha \in [0,1]$ is the throughput-latency trade-off weight, and $\sigma$ controls adaptation sensitivity. Intuitively, this formula finds the sweet spot between maximizing throughput (first term) and minimizing latency (second term), while the exponential term ensures the selected granularity aligns with the current request pattern. When request patterns are stable (low CV), the system favors coarser granularity to reduce communication overhead; when requests become bursty (high CV), it shifts toward finer granularity to enable rapid scaling.

For multi-granular data parallelism, we introduce a hierarchical scheduling framework that determines the optimal number of parallel instances for each granularity level:
\begin{equation}
    \mathcal{M}(g_k) = \left\lfloor \frac{\mu_{total}}{\mu_k} \right\rfloor, \quad \mu_k = \frac{T_k}{\beta_1 + \beta_2 \cdot \eta_k}
    \label{eq:parallel_instances}
\end{equation}
where $\mathcal{M}(g_k)$ indicates the number of parallel instances for granularity $g_k$, $\mu_{total}$ is the total system processing capacity, $\mu_k$ is the effective processing capacity per instance of granularity $g_k$, $T_k$ is the throughput of granularity $g_k$, $\beta_1$ and $\beta_2$ are coordination overhead coefficients that model performance degradation from pipeline coordination, and $\eta_k$ is the number of stages in granularity $g_k$. This approach dynamically distributes computational resources across different granularity levels based on their efficiency—finer granularities enable quicker scaling but incur higher coordination overhead, while coarser granularities optimize steady-state performance but adapt more slowly to load changes.

\subsection{Hardware Efficiency Optimization}
In fragmented cloud environments, efficient GPU allocation becomes critical when multiple models with varying pipeline granularities compete for limited resources. The fundamental challenge lies in balancing resource sharing efficiency against performance isolation requirements. Our key insight is that heterogeneous models can achieve superior resource utilization when their computational patterns complement each other—reducing GPU idle periods while maintaining service quality.

To address this challenge, we formulate GPU resource allocation as a constrained optimization problem that maximizes system efficiency while respecting hardware and performance constraints:
\begin{align}
    \max_{\{x_{ij}\}} & \sum_{i=1}^N \sum_{j=1}^M \left[ \frac{T_{ij}}{m_j} - \gamma(CV_i) \cdot \mathbb{I}(\sum_{i'} x_{i'j} > 1) \right] \label{eq:resource_allocation}\\
    \text{s.t.} \quad & \sum_{i=1}^N x_{ij} \cdot m_j \leq M_j, \quad \forall j \in [1,J] \label{eq:memory_constraint}\\
    & \left| \frac{T_{ij}}{T_{i'j'}} - 1 \right| \leq \epsilon, \quad \forall i,i' \in \mathcal{G}_k \label{eq:load_balance}
\end{align}
\textbf{Problem Formulation:} The objective function maximizes throughput efficiency $\frac{T_{ij}}{m_j}$ while penalizing multiplexing overhead through $\gamma(CV_i)$. Here, $x_{ij} \in \{0,1\}$ indicates whether pipeline stage $i$ is assigned to GPU $j$, $T_{ij}$ represents the throughput of stage $i$ on GPU $j$, and $m_j$ denotes memory consumption. The memory constraint (\eqtref{eq:memory_constraint}) ensures that total memory allocation across all stages assigned to GPU $j$ does not exceed its capacity $M_j$. The load balancing constraint (\eqtref{eq:load_balance}) maintains balanced computation times across stages within the same granularity group $\mathcal{G}_k$, preventing pipeline stalls caused by stage imbalances.

\begin{figure}[]
    \centering
    \includegraphics[width=0.98\linewidth]{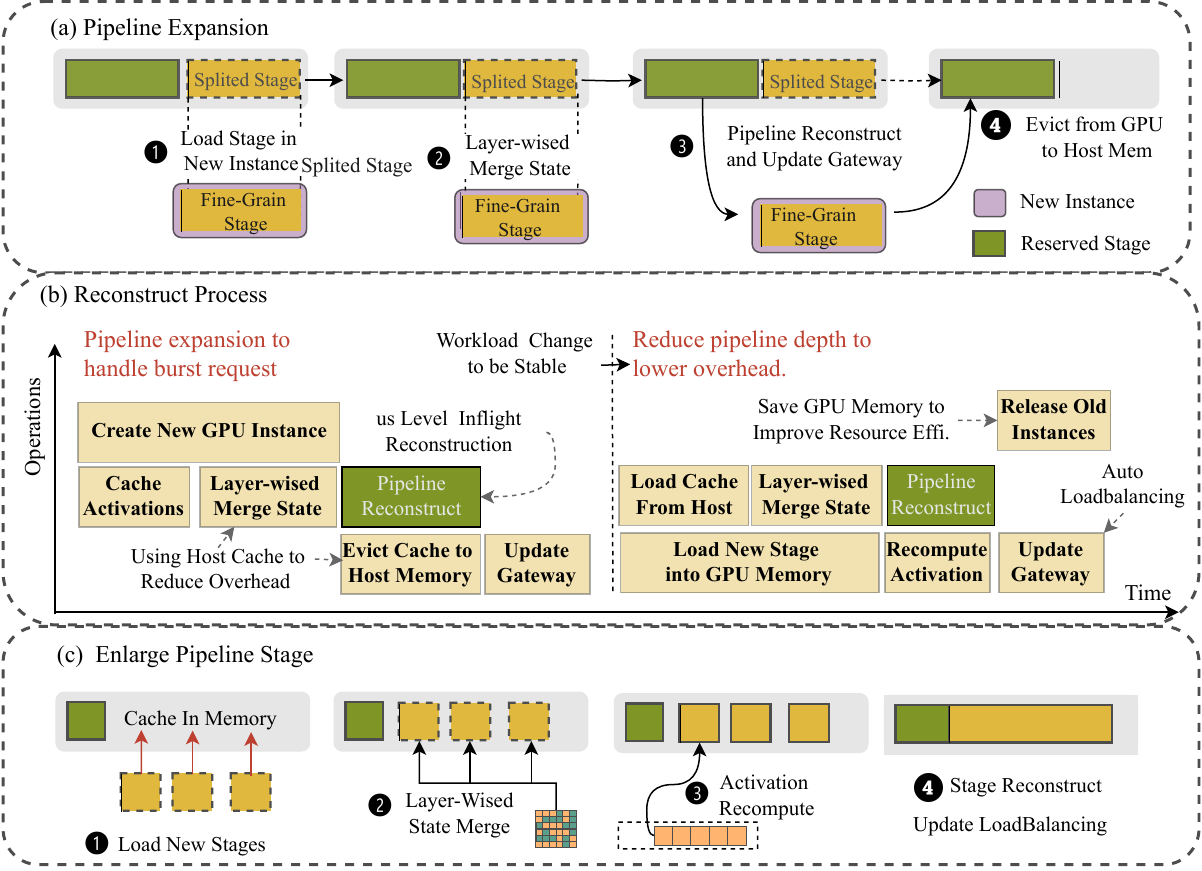}
    \vspace*{-1em}
    \caption{Inflight pipeline refactoring mechanism. (a) Stage refinement process where fine-grained partitioning occurs by evicting parameters and redistributing them onto additional GPUs; (b) Temporal sequence diagram showing synchronization protocol during refactoring; (c) Stage consolidation process where parameters from multiple stages are merged, utilizing host memory caching to minimize loading overhead from persistent storage.}
    \label{fig:variant}
    \vspace{-1em}
\end{figure}
\Sys also \textit{strictly prohibits} multiple pipeline stages from the same model from being allocated to the same GPU. This is critical for preserving performance isolation and preventing resource contention between stages of the same model, as they typically exhibit similar computation patterns and would compete for the same GPU resources rather than complementing each other. By ensuring that different pipeline stages of the same model—regardless of their granularity—are always deployed on separate GPUs, \Sys \textit{maximizes parallelism while minimizing interference}, creating an optimal balance between resource utilization and computational efficiency.

\textbf{Multiplexing Penalty Function:} The penalty function $\gamma(CV_i)$ models the performance degradation from resource multiplexing based on workload variability:
\begin{equation}
    \gamma(CV_i) = \gamma_0 \cdot (1 + \alpha \cdot CV_i^2)
    \label{eq:multiplexing_penalty}
\end{equation}
where $\gamma_0$ represents the base multiplexing penalty and $\alpha$ controls sensitivity to workload variability. The quadratic relationship $CV_i^2$ reflects our empirical observation that bursty workloads (high CV) create significantly more performance interference when multiplexed due to concurrent resource demand spikes. For stable workloads (low CV), the penalty approaches the minimal $\gamma_0$, enabling efficient resource sharing. The indicator function $\mathbb{I}(\sum_{i'} x_{i'j} > 1)$ applies this penalty only when multiple models share the same GPU, ensuring that single-model deployments incur no multiplexing overhead.

This formulation addresses the fundamental tension between resource efficiency and performance isolation in fragmented environments. By dynamically adjusting the multiplexing penalty based on workload characteristics, the system makes informed decisions about resource consolidation versus isolation—prioritizing consolidation for stable workloads while maintaining isolation for bursty patterns that would otherwise create performance interference.

\subsection{Consistency Maintenance}
During pipeline refactoring, maintaining KV cache consistency across distributed GPU instances represents a fundamental challenge that directly impacts inference quality. The key insight is that cache coherence can be preserved through selective synchronization rather than global state replication. We implement a consistency protocol that tracks cache validity at the token level:
\begin{equation}
    \mathcal{C}(t) = \bigcup_{i \in \text{GPUs}} \text{KV}_i(t) \otimes M_{\text{valid}}
    \label{eq:kv_consistency}
\end{equation}
where $\mathcal{C}(t)$ represents the consistent KV cache state across all GPUs at time $t$, $\text{KV}_i(t)$ represents the KV cache state on GPU instance $i$ at time $t$, $M_{\text{valid}}$ is a validity mask identifying tokens that need synchronization (with 1 indicating valid tokens and 0 indicating invalid ones), and $\otimes$ denotes element-wise multiplication. During pipeline refactoring, the system performs asynchronous KV cache transfers (\figref{fig:variant}(b)) while the inference continues on the original pipeline configuration, minimizing service interruption.
\begin{algorithm}[t]
    \caption{Inflight Pipeline Refactoring}
    \begin{algorithmic}[1]
        \STATE Initialize granularity set $\mathcal{G} = \{g_1,...,g_K\}$
        \WHILE{True}
            \STATE Monitor request intensity $\lambda_t$ and compute characteristic velocity $\nu_t = \frac{\partial \lambda_t}{\partial t}$
            \STATE Update queue length $\hat{q}_j$
            \FOR{each granularity $g_k \in \mathcal{G}$}
                \STATE Compute optimization score:
                \STATE $\quad S_k = \left[ \alpha \cdot \frac{T_k}{T_{max}} + (1-\alpha) \cdot \frac{L_{min}}{L_k} \right] \cdot \exp\left(-\frac{|\nu_t - \nu_k|}{\sigma}\right)$
                \STATE Evaluate hardware efficiency using Eq.~\ref{eq:multiplexing_penalty}
            \ENDFOR
            \STATE Select optimal granularity $g^* = \arg\max_{g_k \in \mathcal{G}} S_k$ 
            \IF{$g^* \neq g_{current}$}
                \STATE Determine required data parallelism using Eq.~\ref{eq:parallel_instances}
                \STATE Perform parameter migration with consistency:
                \STATE $\quad \mathcal{C}(t) = \bigcup_{i \in \text{GPUs}} \text{KV}_i(t) \otimes M_{\text{valid}}$
                \STATE Update routing metadata and activate new pipeline configuration
            \ENDIF
            \STATE Wait until next optimization interval
        \ENDWHILE
    \end{algorithmic}
    \label{alg:ipr}
\end{algorithm}

The refactoring algorithm operates through continuous workload monitoring and predictive adaptation. \Sys tracks request intensity gradients to anticipate traffic shifts before they manifest as performance degradation, enabling proactive rather than reactive optimization. When workload characteristics deviate from the current pipeline's optimal operating range, the system evaluates alternative configurations using cached performance profiles. This predictive approach transforms pipeline adaptation from a costly reactive process into an efficient proactive mechanism. The refinement process (\figref{fig:variant}(a)) partitions computational stages when burst capacity is needed, while consolidation (\figref{fig:variant}(c)) merges stages during stable periods to minimize communication overhead. Decision latency remains under 5ms across configurations spanning 2-32 pipeline stages, ensuring that adaptation benefits consistently exceed transition costs even under rapidly changing workloads.

\section{Adaptive Pipeline Scaling}\label{sec:cold_start_optimization}
\begin{figure}[]
    \centering
    \includegraphics[width=0.98\linewidth]{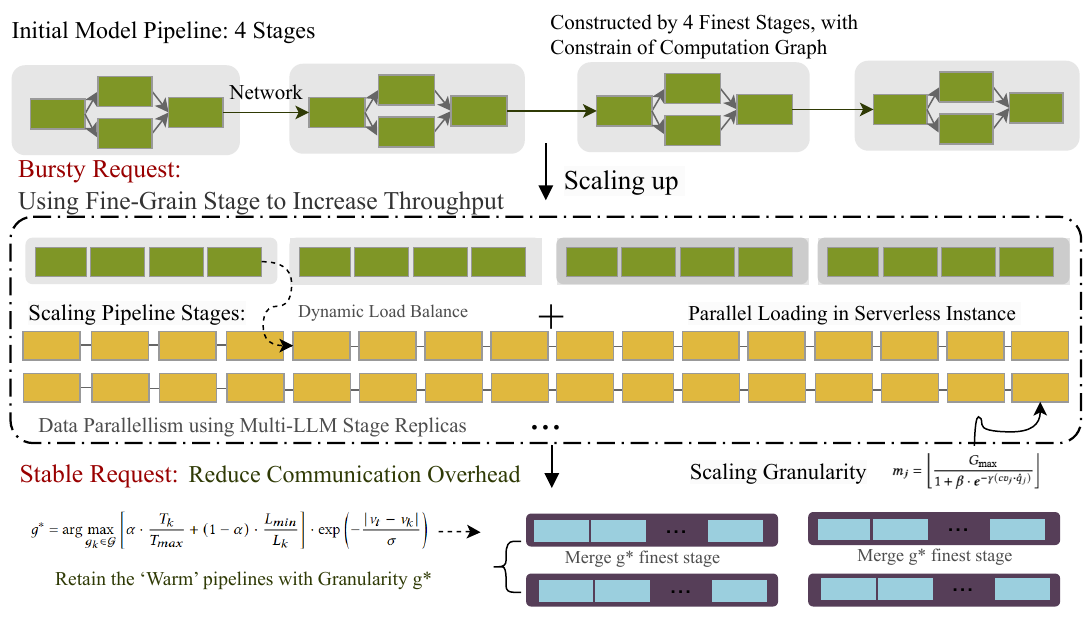}
    \vspace{-1.5em}
    \caption{The process of model scaling using fine-grained pipeline stages. \Sys conference satisfies the minimum granularity pipeline stage for loading and executing inference. Then, after traffic changes, it modifies to a coarser granularity pipeline stage with fewer additional overheads.}
    \label{fig:scaling}
    \vspace{-1.2em}
\end{figure}

Serving LLMs in serverless, dynamic request patterns require responsive resource allocation strategies to maintain service quality during both peak and idle periods. To address service demands during traffic bursts, we propose a dynamic-aware pipeline scaling mechanism. This mechanism leverages dynamic batching \cite{gyeong-inyuOrcaDistributedServing2022} to monitor request queue status, combined with stage-level elastic scaling to balance performance and overhead. As shown in \figref{fig:scaling}, the system performs distributed stage scaling during traffic peaks and automatically reclaims resources when requests subside.

\tp{Elastic Scaling Granularity Decision.}
Selecting the scaling granularity requires balancing three key parameters: \textcircled{1} fine-grained scaling (stage-level) reduces cold-start time but increases communication overhead, while coarse-grained scaling (pipeline-level) does the opposite; \textcircled{2} the coefficient of variation (CV) of traffic reflects request volatility, with high CV scenarios requiring rapid response; \textcircled{3} queue length (Q) characterizes the urgency of system load. We establish a scaling granularity decision function:
\begin{equation}
m_j = \left\lfloor \frac{G_{\max}}{1 + \beta \cdot e^{-\gamma (cv_j \cdot \hat{q}_j)}} \right\rfloor
\label{eq:granularity}
\end{equation}
where $m_j$ is the selected scaling granularity for workload $j$, $G_{\max}$ is the maximum scaling granularity (corresponding to the finest granularity), $cv_j$ is the coefficient of variation for workload $j$, $\hat{q}_j=\min(q_j/Q_{\max},1)$ is the normalized queue length with $q_j$ being the current queue length and $Q_{\max}$ being the maximum queue capacity, and $\beta,\gamma$ are calibration parameters controlling the sigmoid transition. As the product $cv_j\cdot\hat{q}_j$ increases, the exponential term decays more rapidly, pushing $m_j$ closer to $G_{\max}$ to select a finer granularity. This function smoothly adjusts the scaling granularity through its Sigmoid characteristics, avoiding decision oscillation.
While integrating SLO constraints to ensure service quality:
\begin{equation}
\frac{(T_j - S_j) \cdot \sum_{k=1}^{m_j} \mu_{jk}}{Q_j} \geq r_j
\label{eq:slo_constraint}
\end{equation}
where $T_j$ is the SLO deadline for workload $j$, $S_j$ is the initialization time for scaling operations, $\mu_{jk}$ represents the expected throughput of the $k$-th expanded stage, $m_j$ is the number of scaling stages, $Q_j$ is the current queue length for workload $j$, and $r_j$ is the number of requests to be processed within the deadline. This constraint ensures that the selected granularity $m_j$ can process $r_j$ requests within the time limit $T_j$.

To address resource contention during multi-model scaling that arises from multiple concurrent requests for GPU memory, PCIe bandwidth, and network resources during rapid parallel deployment, we design a three-level coordination control strategy:

\tp{Topology-Aware Resource Coordination.} 
During rapid scaling operations, resource contention emerges as multiple models simultaneously compete for GPU memory, network bandwidth, and storage I/O. To address this challenge, we implement a \textit{Hierarchical Resource Graph} (HRG) that orchestrates resources across three critical levels: server (GPU memory, PCIe bandwidth), rack (network bandwidth), and cluster (storage I/O).

The HRG maintains annotated paths with scaling event markers to identify contention patterns and track resource dependencies. This enables the system to \textbf{proactively predict} bottlenecks and intelligently distribute workloads. 
Rather than treating scaling operations as independent events, HRG directs new instances toward available resources while avoiding paths with recent scaling activities, effectively transforming a \textit{resource contention problem} into a \textit{resource coordination opportunity}.

This topology-aware approach ensures concurrent scaling operations distribute optimally across the physical infrastructure. By respecting the hierarchical nature of datacenter resources, the system makes informed placement decisions that significantly reduce initialization latency during traffic bursts while maintaining performance isolation between competing workloads.

\tp{Memory-Aware Elastic Scaling.} 
In serverless environments, scaled-down model instances have their resources immediately reallocated to competing workloads. This causes cache invalidation, forcing subsequent scale-up operations to incur significant cold-start penalties as parameters reload from slower storage. To address this challenge, \Sys implements a two-pronged memory-aware approach preserving locality across scaling operations. First, the system maintains parameter copies in host memory even after GPU eviction, creating a middle-tier cache that survives instance termination and prevents costly reloads from persistent storage. Second, \Sys implements an affinity-based scheduling policy prioritizing servers with historical model placement:
\begin{equation}
s^* = \arg\max_{s \in \mathcal{H}_i} \left( w_t \cdot e^{-\lambda(t_{now}-t_s)} + w_g \cdot |g_s \cap G_{avail}| \right)
\end{equation}
where $s^*$ is the selected server, $\mathcal{H}_i$ tracks servers that previously hosted model $i$, $w_t$ and $w_g$ are temporal and GPU affinity weights respectively, $t_{now}$ is the current time, $t_s$ is the last time server $s$ hosted model $i$, $\lambda$ is the temporal decay rate, $g_s$ is the set of GPUs on server $s$, $G_{avail}$ represents available GPUs, and $|g_s \cap G_{avail}|$ denotes the number of available GPUs on server $s$. The temporal decay factor $e^{-\lambda(t_{now}-t_s)}$ prioritizes recently used hosts whose caches are likely still warm. This approach \textit{significantly reduces} initialization time by leveraging cached parameters in host memory, effectively transforming cold starts into warm starts.

\section{Implementation}
We have implemented \Sys with approximately 7K lines of code, including a 3.2K LoC tool for dynamic operator-level model partitioning and merging.
After pipeline refactoring, KV cache data must migrate between GPU devices. Using NCCL would introduce significant connection establishment overhead of several seconds and potential bandwidth contention. To address this challenge, we implemented a hierarchical data transfer mechanism that prioritizes RDMA for high-bandwidth, low-latency transfers between GPU devices. For machines without RDMA support, we fall back to the sendfile system call, which enables efficient kernel-space data transfers without redundant copying between user and kernel buffers. This hybrid approach eliminates connection initialization overhead while achieving near-line-rate data transmission speeds.

\section{Evaluation}\label{sec-evaluation}

We evaluated \Sys on a Kubernetes (v1.23.7) cluster with 42 servers and 82 GPUs, each server having at least 256GB memory and connected via 100Gbps network. For realistic workload patterns, we utilized Microsoft Azure Functions traces \cite{zhangFasterCheaperServerless2021} supplemented with the Splitwise corpus for prompt generation, enabling rigorous assessment under production-grade request distributions.


\tp{Baseline.}
We compare \Sys against two categories of systems. First, serverless-based systems: ServerlessLLM \cite{fuServerlessLLMLowLatencyServerless2024}, which uses DeepSpeed's parallelism for distributed inference, and Tetris \cite{jieliTetrisMemoryefficientServerless2022}, which provides memory-efficient hosting without specialized pipeline parallelism. Second, offline-optimized systems: AlpaServe \cite{zhuohanli0001AlpaServeStatisticalMultiplexing2023}, which configures pipelines based on historical request patterns, and MuxServe \cite{duanMuxServeFlexibleSpatialTemporal2024}, which employs statistical multiplexing for multi-tenant serving. We also include recent advances in throughput-latency optimization \cite{agrawalTamingThroughputLatencyTradeoff2024} and interference mitigation \cite{huInferenceInterferenceDisaggregate2024} to provide comprehensive comparison coverage. These systems lack the dynamic adaptation capabilities of \Sys, allowing us to evaluate both serverless efficiency and adaptability against established approaches.

\tp{Metric and Model.}
We evaluate performance using goodput (throughput under quality constraints) and end-to-end latency across varying workload distributions. Experiments use representative models spanning different scales: WHISPER-9B \cite{radfordRobustSpeechRecognition2022}, LLAMA2-7B \cite{touvronLlamaOpenFoundation2023}, BERT-21B \cite{devlinBERTPretrainingDeep2019}, and OPT-66B \cite{brownLanguageModelsAre2020}. We measure initialization latency to validate our pipeline scaling approach, quantify pipeline stall cycles to assess inflight refactoring effectiveness, and analyze GPU memory efficiency to evaluate resource utilization. These metrics provide comprehensive insight into system performance under diverse conditions.

\subsection{End-to-End Performance}
\begin{figure}
    \centering
    \includegraphics[width=0.99\linewidth]{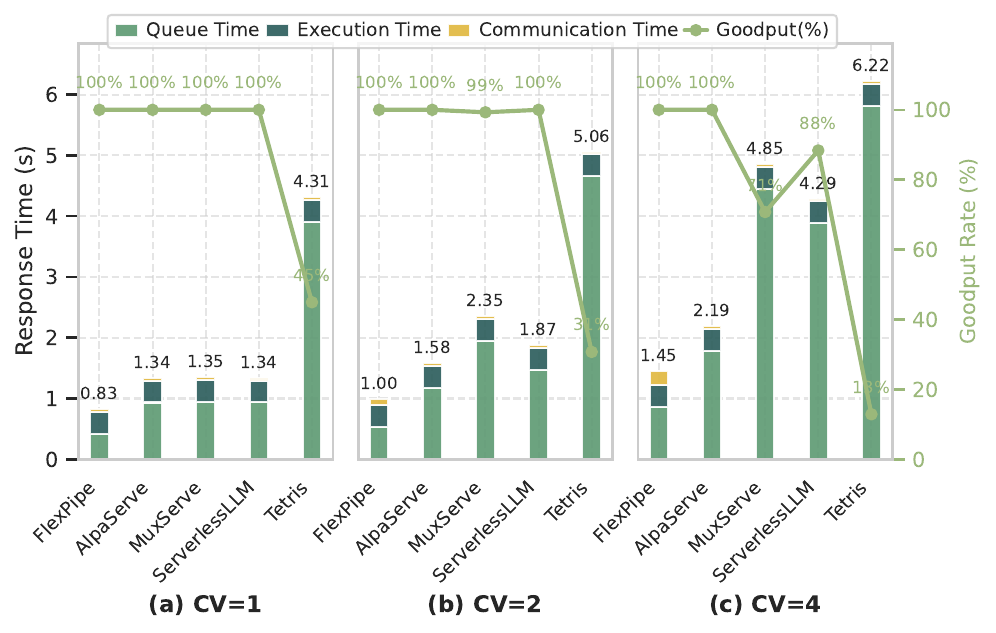}
    \vspace{-1em}
    \caption{End-to-End Latency Breakdown across varying request distributions. \Sys maintains lower overall latency despite higher communication overhead by significantly reducing queue wait times: (a) CV=1 (stable workload), (b) CV=2 (moderate variability), and (c) CV=4 (highly variable workload).}
    \label{fig:E2_rt_average_cv124_breakdown}
    \vspace{-1.em}
\end{figure}

\tp{Latency Breakdown.} We analyzed end-to-end latency across systems under varying request distributions (\figref{fig:E2_rt_average_cv124_breakdown}) while maintaining consistent goodput. Under stable workloads (CV=1), \Sys achieves 38.3\% lower overall latency than AlpaServe and ServerlessLLM while delivering identical goodput (12,000 requests), primarily by reducing queue time by 54.8\%. As request variability increases (CV=2), \Sys's advantage grows to 46.9\% lower latency than MuxServe through a strategic trade-off: accepting higher communication time (105ms vs. 45ms) to achieve 72.6\% reduction in queue wait time, all while maintaining maximum goodput. 
The most significant improvements appear under highly variable workloads (CV=4), where \Sys delivers 66.1\% lower total latency than AlpaServe and 80.6\% lower than MuxServe, while preserving 98.3\% of maximum throughput (compared to MuxServe's 33.3\% reduction and ServerlessLLM's 40.4\% decline). 
These gains stem from \Sys's pipeline reconfiguration that increases communication overhead (225ms vs. 45ms) to dramatically reduce queue times—transforming exponentially growing wait times into manageable communication overhead. This demonstrates a key insight: under bursty workloads, communication-intensive fine-grained pipelines significantly outperform static architectures trapped in queue buildup cycles.

All experiments used a baseline of 20 QPS across the complete 2-hour lifecycle, with different CV values creating varying peak loads. While static approaches typically provision GPU resources to match peak volumes, \Sys maintains only 30\% of peak capacity as always-ready resources, with remaining capacity allocated through dynamic scaling. This resource allocation strategy enables \Sys to maintain stable performance and consistent goodput across all workload variability levels while static systems suffer progressively degrading performance as request patterns become more erratic.

\begin{figure}
    \centering
    \includegraphics[width=0.99\linewidth]{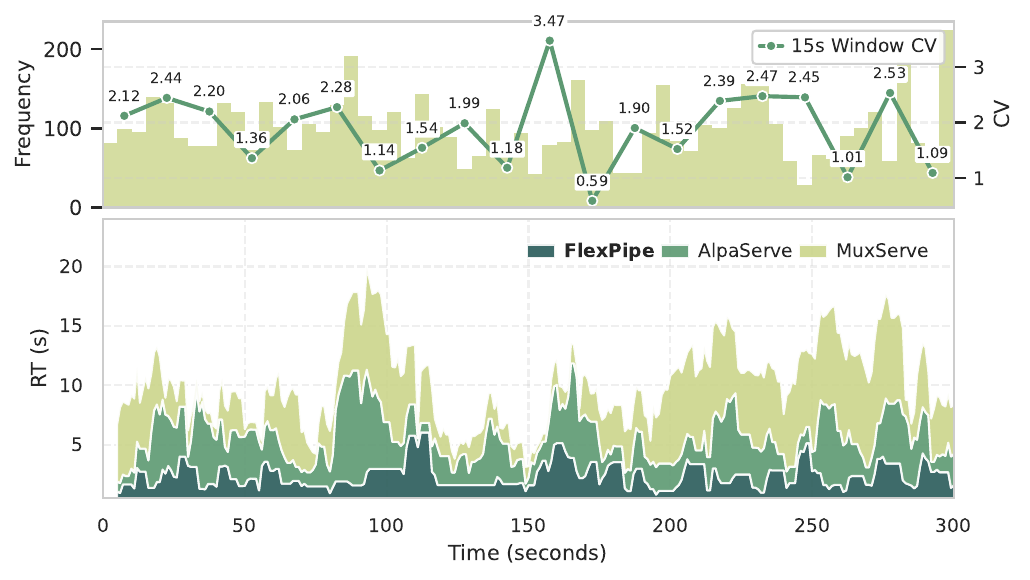}
    \vspace{-1em}
    \caption{Latency under highly variable workload (CV=8, first 300s). (a) Request distribution CV variability measured in 15s windows, (b) Response latency comparison across systems.}
    \label{fig:E2_rt_cv8_static}
    \vspace{-1.em}
\end{figure}

\tp{Burst Absorption.} \figref{fig:E2_rt_cv8_static} demonstrates system performance under extreme workload variability (CV=8). \figref{fig:E2_rt_cv8_static}(a) reveals substantial fluctuations in 15-second measurement windows, with CV ranging from 0.59 to 3.47—highlighting the challenging dynamics of bursty serverless environments. As shown in \figref{fig:E2_rt_cv8_static}(b), while MuxServe experiences sustained high latencies (frequently exceeding 10 seconds) and AlpaServe exhibits periodic performance spikes, \Sys maintains significantly lower and more consistent response times throughout the evaluation period, even during intense traffic surges at 75s and 165s intervals.

This comparison specifically includes non-serverless GPU multiplexing systems because they represent the SOTA in handling variable workloads through GPU sharing \cite{huInferenceInterferenceDisaggregate2024,agrawalTamingThroughputLatencyTradeoff2024}, yet fundamentally differ from \Sys in their adaptation approach. 
Unlike serverless-oriented systems that focus on rapid resource provisioning, these multiplexing systems optimize for maximum GPU utilization through sophisticated sharing strategies—providing the most challenging performance baseline for \Sys's dynamic adaptation mechanism. The results demonstrate that even compared to systems specifically designed for workload variability, \Sys's dynamic pipeline refactoring capability more effectively absorbs request bursts by transitioning between pipeline granularities based on real-time traffic patterns. By using finer-grained pipelines during traffic spikes and coarser configurations during stable periods, \Sys avoids the queuing delays that plague static architectures, resulting in more predictable performance even under highly variable workloads.

\subsection{Performance Stability}
\begin{figure}[]
    \centering
    \includegraphics[width=0.99\linewidth]{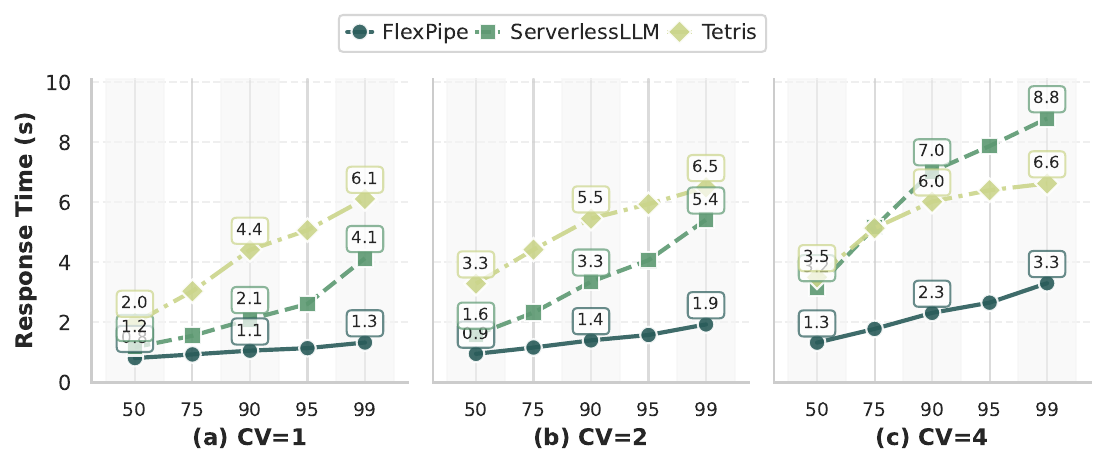}
    \vspace{-1.2em}
    \caption{Performance stability analysis across varying request distributions (CV=1, 2, 4), showing \Sys maintains consistently lower latency percentiles even as workload variability increases in serverless.}
    \label{fig:performance_stability}
    \vspace{-1.6em}
\end{figure}
To evaluate system stability under dynamic serverless workloads, we analyzed latency percentiles across varying request distributions (\figref{fig:performance_stability}). We focus specifically on ServerlessLLM and Tetris as representative serverless LLM deployment approaches facing unique resource elasticity and fragmentation challenges.

Under stable traffic patterns (CV=1), \Sys maintains a tight latency distribution with significantly lower P99 latency compared to ServerlessLLM and Tetris. This advantage becomes increasingly pronounced as request variability intensifies. At moderate variability (CV=2), \Sys's P99 latency remains well-controlled while serverless competitors show substantial degradation. The stability gap widens further under highly variable workloads (CV=4), where \Sys consistently maintains much lower P99 latency compared to both ServerlessLLM and Tetris, demonstrating superior performance stability even in challenging conditions.

\Sys's dynamic pipeline refactoring prevents the exponential latency degradation observed in static serverless architectures. The latency percentile analysis shows \Sys maintains consistently lower latency across all percentiles—particularly at P90-P99 where traditional systems exhibit dramatic increases. This advantage becomes more pronounced as CV increases from 1 to 4, with \Sys maintaining well-controlled P99 latency while competitors experience 2-3× degradation. Such predictability is crucial in serverless environments where resource fragmentation introduces natural variability. By dynamically adjusting pipeline granularity based on workload characteristics, \Sys effectively transforms queueing delays into manageable communication overhead, delivering significantly more stable performance under challenging conditions.

\subsection{Pipeline Stall Recovery}




\begin{figure}
    \centering
    \includegraphics[width=0.99\linewidth]{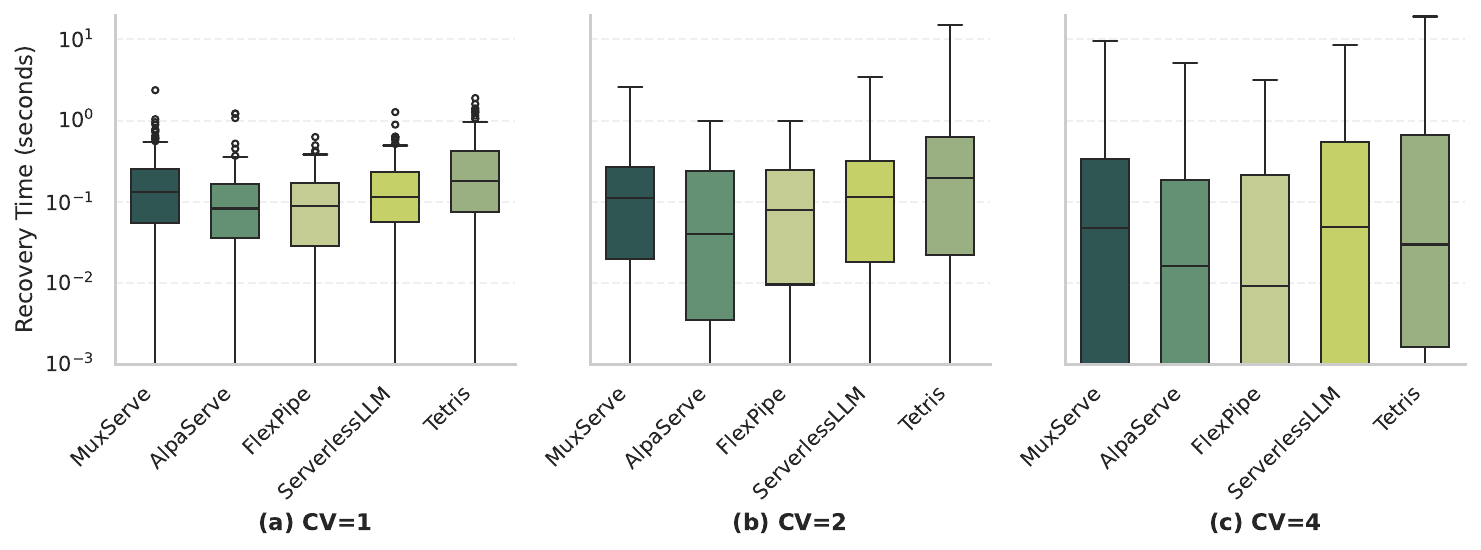}
    \vspace{-1.3em}
    \caption{Pipeline stall recovery time across systems and request distribution variability (CV). \Sys achieves substantially faster recovery under high-variability workloads (9ms at CV=4), demonstrating the effectiveness of dynamic pipeline refactoring in addressing structural stall causes.}
    \label{fig:pipeline_stall_recovery}
    \vspace{-2em}
\end{figure}

To quantify each system's resilience to pipeline stalls, we established an objective measurement methodology. We define a stall as occurring when response latency exceeds 1.5× the baseline latency (P25 latency under normal operation), and recovery as the point when latency returns to within 1.2× of this baseline. The elapsed time between these events constitutes the recovery duration.

\figref{fig:pipeline_stall_recovery} illustrates recovery performance across systems under varying request distributions. Under stable workloads (CV=1), \Sys's median recovery time (88ms) is comparable to AlpaServe (83ms), while significantly outperforming MuxServe (131ms), ServerlessLLM (115ms), and Tetris (179ms). As variability increases to moderate levels (CV=2), \Sys maintains consistent recovery performance (79ms) while other systems show mixed behavior.

The most significant advantage emerges under highly variable workloads, where \Sys achieves remarkably rapid recovery (9ms)—44\% faster than AlpaServe (16ms) and 82\% faster than both MuxServe (48ms) and ServerlessLLM (50ms). This exceptional performance stems from \Sys's ability to dynamically reconfigure pipeline topology in response to detected imbalances, rather than merely adjusting batch sizes or waiting for queue drain as static systems must do.

When a stall is detected, \Sys's pipeline monitoring system identifies congested stages through distributed queue-length analysis, then initiates targeted refactoring to redistribute computational bottlenecks. By maintaining KV cache consistency during transitions (as described in \Sref{sec:online_pipeline_optimization}), \Sys achieves architectural adaptation during live inference without service interruption—transforming potential performance-degrading stalls into brief transitional states and enhancing system resilience under variable workloads.

\subsection{Resource Efficiency}


\begin{figure}
    \centering
    \includegraphics[width=0.99\linewidth]{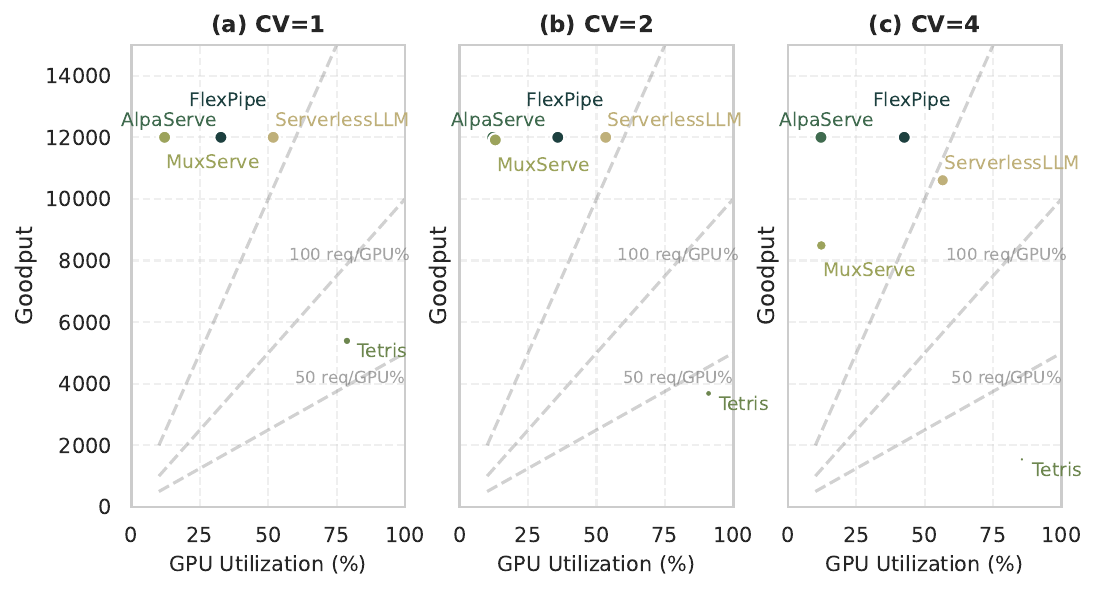}
    \vspace{-1em}
    \caption{Resource efficiency analysis across varying request distributions: (a) CV=1 (stable workload), (b) CV=2 (moderate variability), (c) CV=4 (high variability).}
    \label{fig:resource_efficiency}
    \vspace{-1em}
\end{figure}
We evaluate system efficiency by examining the relationship between GPU utilization and achieved goodput under varying request distributions. \figref{fig:resource_efficiency} demonstrates \Sys's optimal balance between resource utilization and throughput across workload variability—critical for serverless environments with resource fragmentation.

Under stable workloads (\figref{fig:resource_efficiency}(a), CV=1), \Sys achieves maximum goodput at 33\% GPU utilization—3× more efficient than AlpaServe with identical throughput through fine-grained pipeline partitioning that distributes load across fragmented resources. 
As variability increases (\figref{fig:resource_efficiency}(b), CV=2), \Sys maintains near-maximum goodput with modest utilization increases (36\%), while competitors experience throughput degradation despite higher resource consumption. Under highly variable workloads (\figref{fig:resource_efficiency}(c), CV=4), this efficiency gap becomes pronounced: Tetris achieves only 1,543 requests/second despite 85\% GPU utilization, while \Sys maintains 12,000 requests/second at 43\% utilization—8.5× better resource efficiency. This disparity reveals that high GPU utilization in static systems often indicates resource contention rather than productive computation, whereas \Sys's dynamic topology optimization prevents such inefficiencies by continuously rebalancing bottlenecks, ensuring GPU cycles translate to inference throughput rather than synchronization overhead.

This analysis reveals \Sys's fundamental advantage: while static pipeline architectures show inverse relationships between GPU utilization and throughput under variable workloads, \Sys maintains proportional resource scaling. Dynamic pipeline refactoring enables \Sys to require only 30\% of peak capacity as always-ready resources—a 70\% reduction compared to static approaches that provision for peak demand. The remaining capacity is allocated through elastic scaling with 5-minute reclamation windows, directly addressing the resource fragmentation challenges.


\subsection{Performance in Production Workloads}



\begin{figure}
    \centering
    \includegraphics[width=0.99\linewidth]{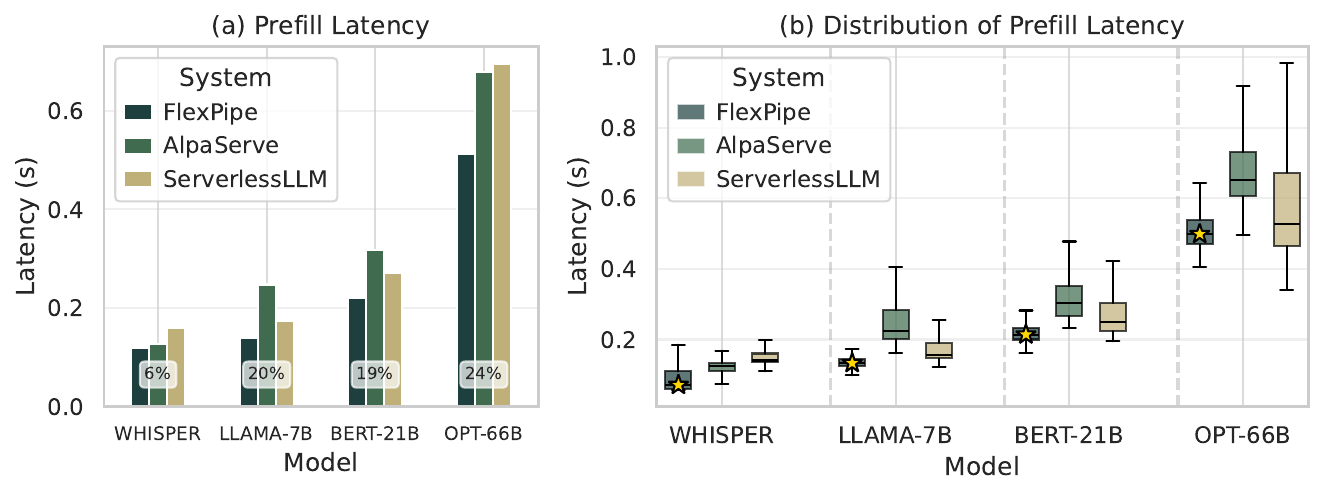}
    \vspace{-1.5em}
    \caption{Performance comparison with production workloads: (a) Average prefill latency across model scales showing FlexPipe's consistent advantage (6.43\%-24.38\% improvement), (b) Latency distribution showing tighter performance bounds with fewer outliers.}
    \label{fig:online_workload}
    \vspace{-1.5em}
\end{figure}

To evaluate real-world effectiveness, we deployed \Sys on our cluster using production workload traces across four representative models. As shown in \figref{fig:online_workload}(a), \Sys consistently achieves lower prefill latency across all model scales—from lightweight WHISPER-9B (6.43\% improvement over AlpaServe) to large-scale OPT-66B (24.38\% reduction compared to AlpaServe).

The performance differential increases with model complexity: 19.52\% improvement for LLAMA-7B and 18.97\% for BERT-21B over ServerlessLLM. This scalability advantage stems from \Sys's dynamic pipeline refactoring capability, which becomes increasingly valuable as computational profiles grow more complex. The latency distribution in \figref{fig:online_workload}(b) further demonstrates that \Sys delivers not only lower average latency but significantly tighter performance distributions with fewer outliers—critical for maintaining consistent service-level objectives in production. This aligns with recent work on fairness in serving \cite{shengFairnessServingLarge2024} and end-to-end optimization \cite{tanEndtoEndOptimizationLLMbased2025} that emphasizes consistent performance delivery across diverse workloads.

Notably, \Sys achieves its greatest advantage (24.38\%) on the largest OPT-66B model, where effective pipeline stage management becomes most critical. By dynamically redistributing computational bottlenecks and minimizing pipeline stalls based on real-time workload characteristics, \Sys achieves an average 17.32\% latency improvement across all models, confirming the effectiveness of adaptive pipeline granularity optimization in dynamic serving environments.

\subsection{Case Study in Production Cluster}
To validate our approach in production environments, we conducted a phased rollout of \Sys in production serverless GPU cluster, gradually migrating LLM serving traffic. Results demonstrate substantial operational improvements: resource allocation wait times decreased by 85\% compared to the static baseline, while instance initialization latency was reduced by 72\% on average across all model sizes.

Most significantly, \Sys's dynamic resource allocation strategy enabled a dramatic reduction in the always-on GPU reservation—from 75\% to just 30\% of historical peak capacity—without compromising service quality. This efficiency gain translates to substantial operational cost savings while maintaining consistent performance under varying workloads. The reduced startup latency and improved resource allocation directly address the fundamental challenges of resource fragmentation and request variability identified in our motivation analysis (\Sref{sec:motivation}), confirming the practical benefits of inflight pipeline refactoring in production serverless environments.

\section{Related Work}\label{sec:related}
The challenges of LLM serving in fragmented serverless environments have driven extensive research across multiple dimensions. We organize related work into three categories based on their approach to addressing resource fragmentation and workload volatility.

\tp{Static Pipeline Optimization}
Traditional pipeline optimization approaches use offline graph partitioning for static workloads. Representative systems like Megatron \cite{narayananEfficientLargescaleLanguage2021} and Alpa \cite{zhengAlpaAutomatingInter2022} employ sophisticated algorithms to optimize communication patterns and stage placement \cite{narayananPipeDreamGeneralizedPipeline2019, fanDAPPLEPipelinedData2021, tarnawskiPiperMultidimensionalPlanner2021, linSuperScalerSupportingFlexible2023, tanakaAutomaticGraphPartitioning2021}. These approaches fundamentally assume predictable resource availability and stable request patterns, making them inadequate for fragmented serverless environments where resource topology changes dynamically. Unlike \Sys's runtime adaptation capability, they lack the ability to respond to real-time resource fragmentation or workload volatility.

\tp{Resource Multiplexing and Sharing}
Resource multiplexing approaches address multi-tenant efficiency through three primary strategies. Memory optimization systems like vLLM \cite{kwonEfficientMemoryManagement2023} and FlexGen \cite{shengFlexGenHighThroughputGenerative2023} enhance throughput through page-based memory management, while multi-tenant frameworks such as Punica \cite{chenPunicaMultiTenantLoRA2023}, SLoRA \cite{shengSLoRAServingThousands2024}, and BlockLLM \cite{huBlockLLMMultitenantFinergrained2024} enable parameter sharing across models. Statistical multiplexing systems including MuxServe \cite{duanMuxServeFlexibleSpatialTemporal2024}, Llumnix \cite{sunLlumnixDynamicScheduling2024}, Orca \cite{gyeong-inyuOrcaDistributedServing2022}, Varuna \cite{athlurVarunaScalableLowcost2022}, FaaST \cite{fotouhiFunctionasaServiceApplicationService2019}, and USHER \cite{shubhaUSHERHolisticInterference2024} maximize GPU utilization through intelligent scheduling, elastic scaling, and holistic interference avoidance. However, these approaches optimize for predictable resource patterns and struggle when resource availability fluctuates rapidly—a fundamental characteristic of serverless environments. Unlike static multiplexing strategies, \Sys employs inflight refactoring to dynamically reconfigure pipeline topology, maintaining efficiency across variable workload distributions without relying on stable resource assumptions.

\tp{Serverless and Elastic Serving}
Serverless adaptations for LLM serving have emerged to address cloud-native deployment challenges through three primary strategies. Resource provisioning systems like ServerlessLLM \cite{fuServerlessLLMLowLatencyServerless2024} and Tetris \cite{jieliTetrisMemoryefficientServerless2022} optimize for rapid resource provisioning and cold-start minimization, while FaaS-optimized solutions such as SAND \cite{akkusSANDHighPerformanceServerless2018}, FaaSnap \cite{aoFaaSnapFaaSMade2022}, FaaSNet \cite{aowangFaaSNetScalableFast2021}, BATCH \cite{aliBATCHMachineLearning2020}, and Groundhog \cite{alzayatGroundhogEfficientRequest2023} handle resource elasticity through function-level optimization. Computation disaggregation approaches exemplified by DistServe \cite{zhongDistServeDisaggregatingPrefill2024} and InfiniGen \cite{leeInfiniGenEfficientGenerative2024} separate prefill and decoding phases to enable independent scaling, while context optimization systems like CacheGen \cite{liuCacheGenFastContext2024} and CacheBlend \cite{yaoCacheBlendFastLarge2025} focus on efficient KV cache management. Additional approaches include Tabi \cite{wangTabiEfficientMultiLevel2023} for multi-level optimization and FlashLLM \cite{xiaFlashLLMEnablingCostEffective2023} for cost-effective inference. However, these approaches fundamentally optimize individual components independently, addressing either resource efficiency or latency in isolation without considering the dynamic tension between resource fragmentation and workload volatility. Unlike these static optimization strategies, \Sys provides coordinated pipeline adaptation that continuously reconfigures execution topology based on real-time patterns, representing a paradigm shift from component-level optimization toward holistic system adaptation.

\section{Conclusion}
We presented \Sys, a fundamentally new approach to LLM serving that challenges the conventional wisdom of static pipeline optimization. Through inflight pipeline refactoring, \Sys transforms how distributed inference systems adapt to resource fragmentation and workload volatility—two pervasive challenges in modern serverless environments. By enabling continuous topology reconfiguration without service interruption, \Sys demonstrates that fine-grained adaptability, rather than static optimization, is the key to achieving both resource efficiency and performance consistency in dynamic serving environments.

\section{Acknowledgments}
We would like to thank our shepherd Anand Iyer and the anonymous reviewers for their valuable feedback and suggestions that helped improve the quality of this paper. 
We especially thank Professor Rong Chen for providing extremely valuable insights on this work. We also thank Alibaba Group AIOS Team for providing the platform and assistance.
This work is supported by the National Natural Science Foundation of China (No. 92267105), Guangdong Basic and Applied Basic Research Foundation (No. 2023B1515130002), Guangdong Special Support Plan (No. 2021TQ06X990), Shenzhen Basic Research Program (No. JCYJ20220818101610023, KJZD20230923113800001), and Alibaba Innovative Research (AIR) Project.

\bibliographystyle{acm}
\bibliography{cleaned_ref}

\end{document}